\documentclass[a4paper,fleqn,usenatbib]{mnras}

\usepackage{Times}

\usepackage[T1]{fontenc}
\usepackage{ae,aecompl}

\usepackage{graphicx}	
\usepackage{amsmath}	
\usepackage{amssymb}	
\usepackage{tablefootnote}

\title[Luminous LAEs at $z\approx6-7$]{Spectroscopic properties of luminous Lyman-$\alpha$ emitters at $z\approx6-7$ and comparison to the Lyman-break population}

\author[J. Matthee et al.]{Jorryt Matthee$^{1}$\thanks{E-mail: matthee@strw.leidenuniv.nl}, David Sobral$^{1,2}$, Behnam Darvish$^{3}$, S\'ergio Santos$^{2}$, 
\newauthor Bahram Mobasher$^{4}$, Ana Paulino-Afonso$^{5,6}$, Huub R\"ottgering$^{1}$, Lara Alegre$^{5,6}$\\
$^{1}$ Leiden Observatory, Leiden University, PO\ Box 9513, NL-2300 RA Leiden, The Netherlands\\
$^{2}$ Department of Physics, Lancaster University, Lancaster, LA1 4YB, UK \\ 
$^{3}$ Cahill Center for Astrophysics, California Institute of Technology, 1216 East California Boulevard, Pasadena, CA 91125, USA \\
$^{4}$ University of California, Riverside, 900 University Ave, Riverside, CA 92521, USA \\
$^{5}$ Instituto de Astrof\'{\i}sica e Ci\^{e}ncias do Espa\c{c}o, Universidade de Lisboa, OAL, Tapada da Ajuda, PT1349-018 Lisboa, Portugal \\
$^{6}$ Departamento de F\'{i}sica, Faculdade de Ci\^{e}ncias, Universidade de Lisboa, Edif\'{i}cio C8, Campo Grande, PT1749-016 Lisbon, Portugal
 \\
}

\date{Accepted 2017 August 8. Received 2017 August 8; in original form 2017 June 20.}

\pubyear{2017}

\begin{document}
\label{firstpage}
\pagerange{\pageref{firstpage}--\pageref{lastpage}}
\maketitle

\begin{abstract}
We present spectroscopic follow-up of candidate luminous Ly$\alpha$ emitters (LAEs) at $z=5.7-6.6$ in the SA22 field with VLT/X-SHOOTER. We confirm two new luminous LAEs at $z=5.676$ (SR6) and $z=6.532$ (VR7), and also present {\it HST} follow-up of both sources. These sources have luminosities L$_{\rm Ly\alpha} \approx 3\times10^{43}$ erg s$^{-1}$, very high rest-frame equivalent widths of EW$_0\gtrsim 200$ {\AA} and narrow Ly$\alpha$ lines (200-340 km s$^{-1}$). VR7 is the most UV-luminous LAE at $z>6.5$, with M$_{1500} = -22.5$, even brighter in the UV than CR7. Besides Ly$\alpha$, we do not detect any other rest-frame UV lines in the spectra of SR6 and VR7, and argue that rest-frame UV lines are easier to observe in bright galaxies with low Ly$\alpha$ equivalent widths. We confirm that Ly$\alpha$ line-widths increase with Ly$\alpha$ luminosity at $z=5.7$, while there are indications that Ly$\alpha$ lines of faint LAEs become broader at $z=6.6$, potentially due to reionisation. We find a large spread of up to 3 dex in UV luminosity for $>L^{\star}$ LAEs, but find that the Ly$\alpha$ luminosity of the brightest LAEs is strongly related to UV luminosity at $z=6.6$. Under basic assumptions, we find that several LAEs at $z\approx6-7$ have Ly$\alpha$ escape fractions $\gtrsim100$ \%, indicating bursty star-formation histories, alternative Ly$\alpha$ production mechanisms, or dust attenuating Ly$\alpha$ emission differently than UV emission. Finally, we present a method to compute $\xi_{ion}$, the production efficiency of ionising photons, and find that LAEs at $z\approx6-7$ have high values of log$_{10}(\xi_{ion}$/Hz erg$^{-1}) \approx 25.51\pm0.09$ that may alleviate the need for high Lyman-Continuum escape fractions required for reionisation.

\end{abstract} 



\section{Introduction}
Observations of galaxies in the early Universe help to constrain the properties of the first stellar populations and black holes and to understand the reionisation process and sources responsible for that. However, because of their high redshift, these galaxies are very faint and their rest-frame spectral features (i.e. UV lines) shift to near-infrared wavelengths. This makes spectroscopic observations challenging and currently limited to the brightest sources. Therefore, it has only been possible to study a few galaxies in detail \citep[e.g.][]{Ouchi2013,Sobral2015,Stark2015_CIII,Stark2015_CIV,Zabl2015}. Most of these galaxies are strong Lyman-$\alpha$ (Ly$\alpha$, $\lambda_{0, vac} = 1215.7$ {\AA}) emitters (LAEs). This is partly by selection, as LAEs can easily be identified with wide-field narrow-band surveys \citep[e.g.][]{Konno2014,Matthee2015} and are easier to follow-up spectroscopically, but also because the fraction of UV-bright galaxies with strong Ly$\alpha$ emission increases with redshift \citep[e.g.][]{Curtis-Lake2012,Stark2016}, such that a large fraction of Lyman-break galaxies at $z\approx5-6$ (after reionisation) are typically also classed as LAEs \citep[e.g.][]{Pentericci2011,Stark2011,Cassata2015}, see e.g. \citet{Dayal2012} for a theoretical perspective. 

Ly$\alpha$ photons undergo resonant scattering by neutral hydrogen resulting in significant uncertainties when using Ly$\alpha$ luminosities to study intrinsic properties of galaxies \citep[e.g.][]{Hayes2015}. The fraction of observed Ly$\alpha$ photons depends on the spatial distribution of neutral hydrogen and the characteristics of the emitter \citep[e.g.][]{Matthee2016,Sobral2016}. Hence, high resolution measurements of the Ly$\alpha$ line-profile and measurements of the extent of Ly$\alpha$ can provide information on the properties of both the inter-stellar medium (ISM) and the circum-galactic medium (CGM) \citep[e.g.][]{Moller1998,Steidel2011,Verhamme2015,Arrigoni2016,GronkeDijkstra2016}. Furthermore, the prevalence of Ly$\alpha$ emitters and the Ly$\alpha$ equivalent width (EW) distribution can be used to study the neutral fraction of the inter-galactic medium (IGM) in the epoch of reionisation \citep[e.g.][]{DijkstraReview,Hutter2014}. 

Several observations of LAEs indicate an increasingly neutral fraction at $z>6.5$: at fixed UV luminosity, the fraction of typical Lyman-break galaxies with strong Ly$\alpha$ emission (observed in a slit) is observed to decrease with redshift \citep[e.g.][]{Pentericci2014,Tilvi2014}; the observed number density of LAEs decreases at $z>6$ \citep[e.g.][]{Konno2014,Matthee2015,Zheng2017} and at fixed central Ly$\alpha$ luminosity, there is more extended Ly$\alpha$ emission around faint LAEs at $z=6.6$ than at $z=5.7$ \citep{Momose2014,Santos2016}. These observations all indicate that a relatively larger fraction of Ly$\alpha$ photons are scattered out of the line of sight at $z>6.5$ than at $z<6.5$. Hence, the galaxies that are still observed with high Ly$\alpha$ luminosities at $z>7$ \citep[e.g.][]{Oesch2015,Zitrin2015,Schmidt2016} are likely the signposts of early ionised bubbles \citep[e.g.][]{Stark2016}.

\cite{Matthee2015} performed a survey of LAEs at $z=6.6$, increasing the available number of bright LAEs that allowed detailed study. Two LAEs from this sample (`CR7' and `MASOSA') have been spectroscopically confirmed in \cite{Sobral2015}. Several more recent wide-area surveys at $z=6.6$ and $z=6.9$ are now also identifying LAEs with similar luminosities \citep[e.g.][]{Hu2016,Shibuya2017,Zheng2017}. CR7 and `Himiko' \citep{Ouchi2009} have been the subject of detailed spectroscopic studies \citep[e.g.][]{Ouchi2013,Sobral2015,Zabl2015,Bowler2016}, which indicate that their ISM is likely metal poor and in high ionisation state. Such ISM conditions are similar to those in LAEs at $z\sim2-3$ \citep[e.g.][]{Song2014,Trainor2015,Hashimoto2016,Nakajima2016,Trainor2016}, although we note that the Ly$\alpha$ luminosities of the latter samples are typically an order of magnitude fainter. In order to obtain a comparison sample to those at $z\sim7$, \cite{Santos2016} undertook a comparable survey at $z=5.7$, just after the end of reionisation. A major limitation is that the nature of the most luminous LAEs is currently unknown. Are they powered by active galactic nuclei (AGN) or star formation? What are their metallicities? 

In this paper, we present follow-up observations of candidate luminous LAEs at $z=5.7$ and $z=6.6$ using VLT/X-SHOOTER, which is a high resolution spectrograph with a wavelength coverage of $\lambda=0.3-2.5\mu$m. We assess the interloper and success fractions and use these to update the number densities of the most luminous LAEs. We present the properties of the Ly$\alpha$ lines, UV continua of newly confirmed luminous LAEs, and constrain rest-frame UV nebular lines. Together with a compilation of spectroscopically confirmed LAEs and Lyman-break galaxies (LBGs) from the literature, we study the evolution of Ly$\alpha$ line-widths between $z=5.7-6.6$ and the relation between Ly$\alpha$ luminosity and UV luminosity. Finally, we explore the ionising properties (such as the production efficiency of ionising photons) using an empirical relation to estimate the Ly$\alpha$ escape fraction \citep[e.g.][]{Sobral2016}.

The initial sample of luminous LAEs at $z=5.7$ and $z=6.6$, the observations and data reduction are presented in \S \ref{sec:sample_observations}. We present the results in \S \ref{sec:results}, which include updated number densities. In \S \ref{sec:properties} we present the properties of newly confirmed LAEs. The properties of the sources are discussed and compared to the more general galaxy population at $z\approx6-7$ in \S \ref{sec:discussion}. This section includes a comparison of their Ly$\alpha$ line-widths (\S \ref{sec:widths}), the UV line-ratios to Ly$\alpha$ (\S \ref{sec:compilation}) and their UV luminosity (\S \ref{sec:uvmags}). We discuss their production efficiency of ionising photons in \S $\ref{sec:xion}$. Finally, we summarise our conclusions in \S \ref{sec:conclusions}. Throughout the paper we use a flat $\Lambda$CDM cosmology with $\Omega_M = 0.3$, $\Omega_{\Lambda} = 0.7$ and H$_0 = 70$ km s$^{-1}$ Mpc$^{-1}$.
 
\section{Sample \& Observations} \label{sec:sample_observations}
\subsection{Sample}
The target sample includes candidate luminous LAEs selected through NB816 and NB921 narrow-band imaging with Subaru/Suprime-Cam in the SA22 field over co-moving volumes of $6.3\times10^6$ Mpc$^3$ and $4.3\times10^6$ Mpc$^3$ at $z=5.7$ and $z=6.6$ as described in \cite{Santos2016} and \cite{Matthee2015}, respectively.\footnote{This sample already included the confirmed LAEs Himiko \citep{Ouchi2013}, MASOSA and CR7 \citep{Sobral2015}. It furthermore includes 14 other spectroscopically confirmed LAEs at $z=6.6$ from \citet{Ouchi2010} and 46 spectroscopic confirmed LAEs at $z=5.7$ from \citet{Ouchi2008}, \citet{Hu2010} and \citet{Mallery2012}.} The data in the SA22 field is very wide-field, yet shallow and single-epoch, and is aimed at identifying the brightest LAEs. Even though the expected number of contaminants and transients is significant, these sources are bright enough to be confirmed (or refuted) in relatively small amounts of telescope time.

\begin{table*}
\centering
\caption{Targeted sample of LAE candidates at $z=5.7-6.6$. Candidates selected in NB816 are from \citet{Santos2016}, while candidates selected in NB921 are from \citet{Matthee2015}. L$_{\rm Ly\alpha}$ is the estimated Ly$\alpha$ luminosity from NB imaging. We also list the observation dates from ESO program ID 097.A-0943, the total on-source exposure times and the telluric standard stars that have been used for flux-calibration. The final column identifies the classification of the targets, with 1 = Ly$\alpha$, 2 = [O{\sc iii}], 3 = transient and 4 = star. Sources that are confirmed spectroscopically as Ly$\alpha$ emitters are shown in bold. We provide flux calibrated reduced spectra of SR6 and VR7 with the published version of the paper. }
\begin{tabular}{lrrrp{2.3cm}rrp{1.1cm}r}
\hline
ID & R.A. & Dec. & L$_{\rm Ly\alpha, NB}$ & Dates & t$_{\rm exp, VIS}$ & t$_{\rm exp, NIR}$ & Telluric & Class  \\ 
& J2000 & J2000 & $10^{43}$ erg s$^{-1}$ & 2016 & ks & ks & & \\ \hline 

SA22-NB816-9442& 22:18:00.68 & +01:04:30.53& 8.4 & 5 Aug & 2.92 & 3.12 & GD153 & 2 \\
SA22-NB816-366911& 22:13:00.92& +00:36:24.17  & 4.1 & 7 and 31 Aug & 5.84 & 6.24 & GD153, EG274& 3 \\ 
SA22-NB816-360178 & 22:12:54.85 & +00:32:54.76 & 3.8 & 3 Sep & 2.92 & 3.12 & GD71& 4 \\ 
SA22-NB816-390412 & 22:15:01.22 & +00:46:24.25  & 3.7 & 28 Aug  & 5.84 & 6.24 & Feige110& 3 \\
\bf SR6 & 22:19:49.76 & +00:48:23.90  &3.4  & 2 Sep & 5.84 & 6.24 & GD71& 1  \\ 
SA22-NB816-508969& 22:21:09.92 & +00:47:19.52  & 3.3 & 3 Sep & 2.92 & 3.12 & GD71& 3 \\ 

\bf VR7  & 22:18:56.36 & +00:08:07.32  & 2.4 & 12, 16 Jun, 12 Jul & 8.76 & 9.36 & GD153& 1 \\  
SA22-NB921-D10845 & 22:18:54.82 & +00:06:24.26  & 1.2 & 14 Jul, 2, 3 Aug & 8.76 & 9.36 & GD153, EG274& 3\\ 
SA22-NB921-W210761 & 22:14:38.63 & +00:56:02.98  & 4.1 & 2 Aug  & 2.92 & 3.12 & EG274 & 3\\ 
SA22-NB921-W219795 & 22:15:29.18 &+00:29:17.90 & 3.8 & 3 Aug & 2.92 & 3.12 & EG274 & 3\\ 
SA22-NB921-W6153 & 22:20:20.79 & +00:17:27.96  & 11.0 & 2 Aug & 2.92 & 3.12 & EG274& 3 \\
SA22-NB921-W209855 & 22:16:05.05 & +00:51:59.23  & 3.8 & 3 Aug& 2.92 & 3.12 &  EG274& 3\\ 

\hline\end{tabular}

\label{tab:targets}
\end{table*}

The initial potential target samples included 6 objects at $z=5.7$ and 21 at $z=6.6$. Before choosing the final targets to follow-up spectroscopically, we investigated the individual exposures, instead of only inspecting the final reduced NB image. Nine sources from the NB921 sample were moving solar-system objects whose position changed by $\approx0.2-0.5 ''$ between individual exposures. The stacked image of these sources then resulted in a slightly extended object. Such extended objects in the NB image resemble confirmed LAEs at $z=6.6$ (e.g. Himiko and CR7), leading to their misidentification as candidates. We note that point-like sources may however still be other types of transients/variables. Six other sources from the NB921 sample have been identified as a detector artefact in a single exposure, which coincides with positive noise peaks in the other exposure. Due to PSF-homogenisation these artefacts were then not identified in our visual inspections of the final stack. These checks were also performed for the NB816 candidates, and were excluded already before the final analysis of \cite{Santos2016}. These issues do not  influence the search for LAE candidates in the fields with deeper coverage (COSMOS and UDS), as those fields have been observed with many more individual exposures. The final selection results in a sample of 6 LAE candidates at $z=5.7$ and 6 at $z=6.6$, all in the SA22 field, see Table $\ref{tab:targets}$.

\subsection{Observations}\label{sec:spectra}
We observed the candidate LAEs with the X-SHOOTER echelle spectrograph, mounted on UT2 of the VLT \citep {Vernet2011}. X-SHOOTER simultaneously takes a high resolution spectrum with a UVB, VIS and a NIR arm, providing a wavelength coverage from 300\,nm to 2480\,nm. 

Observations were done under clear skies with a seeing ranging from 0.7-0.9$''$, using 0.9$''$ slits in the NIR and VIS arm and a slow read-out speed without binning. This leads to a spectral resolution of 1.2\,{\AA} ($R\approx7400$) and 3.6\,{\AA} ($R\approx4000$) in the VIS and NIR arm, respectively. We first acquired a star (with $I$-band magnitudes 16-17 AB) and applied a blind offset to the target. In order to improve the NIR sky subtraction, we use the standard \texttt{AutoNodOnSlit} procedure, which nods between two positions A and B along the slit, offset by 3.5$''$. This is repeated two times in an ABBA pattern. At each position, we take a 730\,s exposure in the VIS arm and four 195\,s exposures in the NIR arm. This results in a total exposure time of 2.92\,ks in VIS and 3.12\,ks in NIR in a single observing block. Several sources have been observed in two or three observing blocks, doubling or tripling the total exposure time, see Table $\ref{tab:targets}$.

\subsection{Data reduction}
Data have been reduced with the recipes from the standard X-SHOOTER pipeline \citep{Modigliani2010}, which includes corrections for the bias from read-out noise (VIS arm) and dark current (NIR arm), sky subtraction and wavelength calibration. Since wavelength calibration is done in air, we convert the wavelengths to vacuum wavelengths following \cite{Morton1991}. The standard stars GD71, GD153, EG274 and Feige110 have been observed with a 5$''$ slit for flux calibration. We use the X-SHOOTER pipeline to combine the exposures from single observing blocks. In the case that a source has been observed with multiple observing blocks, we co-add the frames by weighting the sky background and by correcting for slight positional variations based on the position of the peak of observed Ly$\alpha$ lines. 

\begin{figure*}
\begin{tabular}{cc}
	\includegraphics[width=8.6cm]{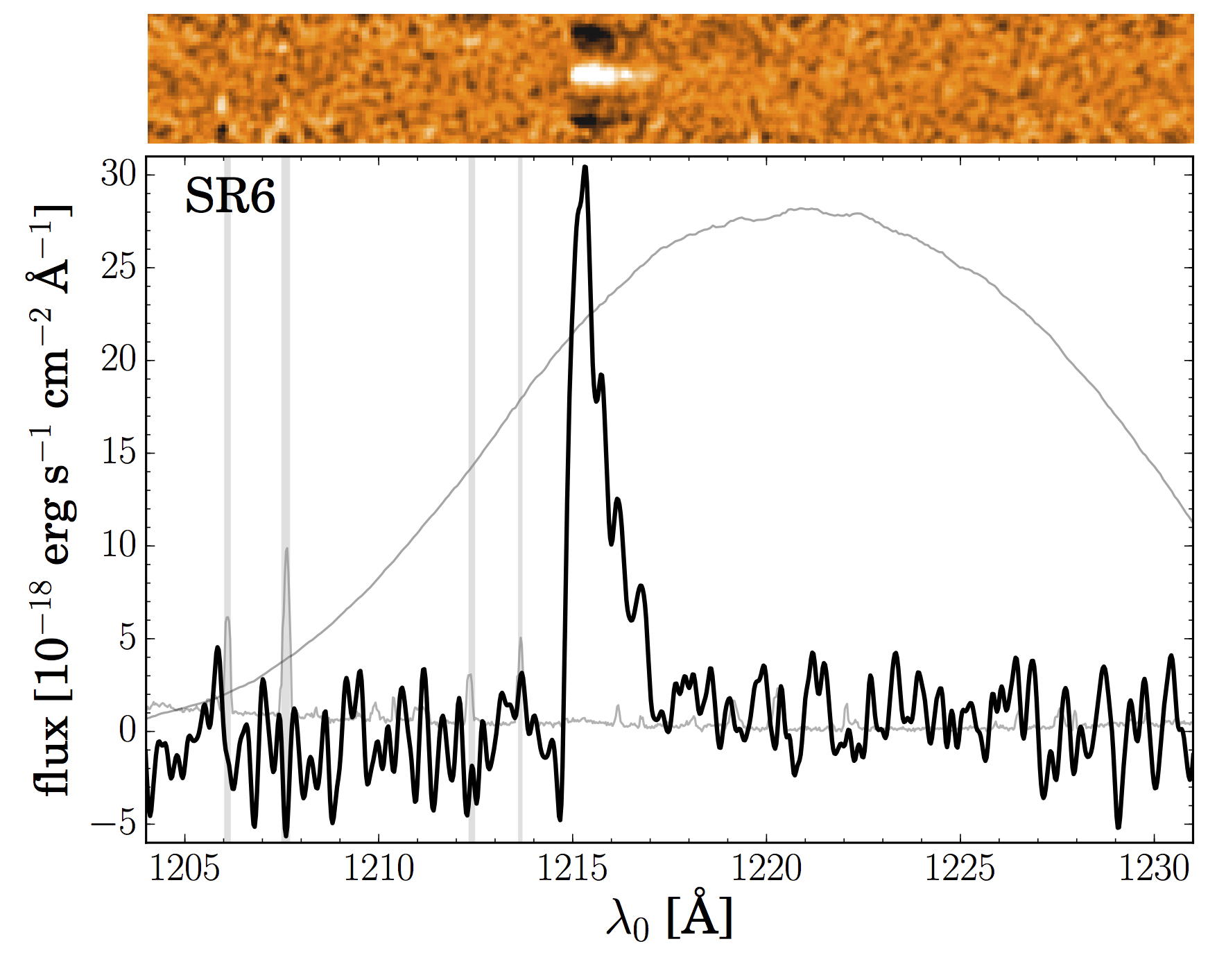} &
		\includegraphics[width=8.6cm]{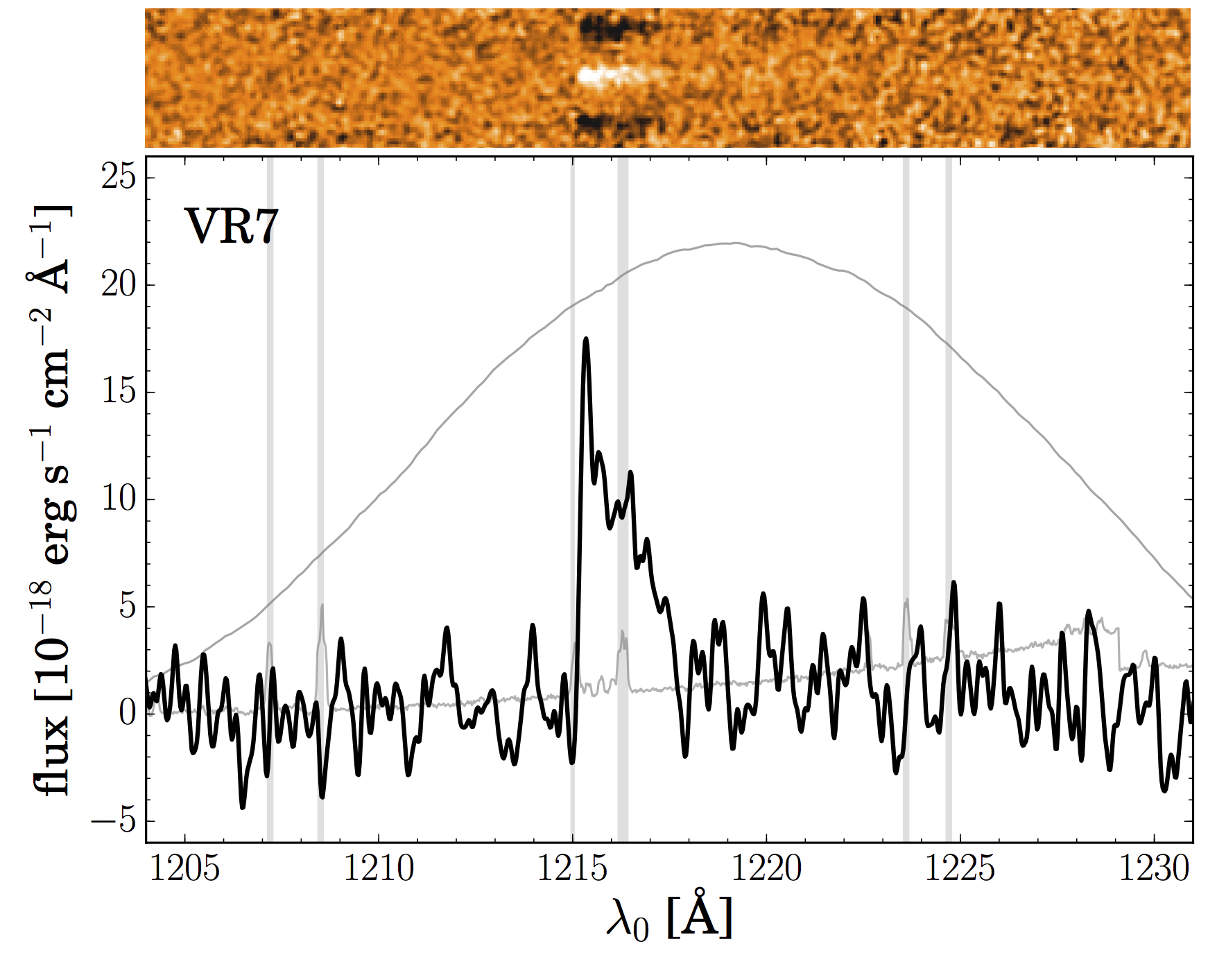} \\
	\end{tabular}
    \caption{Rest-frame X-SHOOTER spectra of the newly confirmed luminous LAEs SR6 at $z=5.676$ and VR7 at $z=6.532$, zoomed in on the Ly$\alpha$ line. In the background we show the NB816/NB921 filter transmission (normalised arbitrarily for visualisation purposes), which illustrates that these sources are detected at $\approx75$ \% and $\approx92$ \% of the peak filter transmission in the original data, respectively. We also illustrate the position of atmospheric OH-lines in the background, based on the noise map provided by the X-SHOOTER pipeline. The Ly$\alpha$ line of VR7 is slightly contaminated by a faint sky-line at $\lambda_0 \approx 1216.3$ {\AA} (observed $\lambda\approx9161$\,{\AA}).
}
\label{fig:spectra}
\end{figure*}
\subsection{Extraction}
We extract 1D spectra in the VIS (NIR) arm by summing the counts in 10 (8) spatial pixels, corresponding to 1.6 (1.68)$''$, along the wavelength direction. These extraction boxes optimise the S/N in confirmed emission-line galaxies in our data-set. Slit losses are estimated by convolving the NB image to the PSF of spectroscopic observations and measuring the fraction of the flux that is retrieved within the slit compared to the flux measured with {\sc Mag-auto}. Typical slit losses are $\approx50-60$ \%. We measure the effective spectral resolution at $\sim 0.9\,\mu$m and $\sim 1.6\,\mu$m by measuring the FWHM of well separated, isolated skylines and find $R=7500$ and $R=4400$, respectively. We note that due to this high resolution, instrumental line-broadening of the emission-lines from the sources discussed in this paper are negligible. 

The line-flux sensitivity is measured as a function of line-width as follows. First, we select the sub-range of wavelengths in the collapsed 1D spectra that are within 30 nm from the targeted wavelength. Then, we measure the flux in 5000 randomly placed positions in this sub-range, with kernels corresponding to the targeted wavelength. We then calculate the noise as the r.m.s. of the 5000 measured fluxes. Note that, in the presence of skylines, this depth is a conservative estimate as it includes flux from skyline residuals, which could increase the noise by a factor $\approx2-3$ depending on the specific wavelength.

\section{Results} \label{sec:results}
\subsection{NB816 targets - candidate LAEs at $\bf z=5.7$}
Out of the six brightest candidate LAEs at $z=5.7$ that we observed with X-SHOOTER, one is reliably confirmed as a Lyman-$\alpha$ emitter, one is identified as [O{\sc iii}] interloper, one is identified as brown dwarf star interloper and three are not detected, indicating that their NB detection was likely due to a transient or variable source, see Table $\ref{tab:targets}$ for a summary.

SA22-NB816-9442 is identified as an [O{\sc iii}] emitter at $z=0.638$. The flux observed in NB816 can be attributed to both the 4959 and 5007 {\AA} lines. We measure a combined line-flux of $0.8\pm0.1\times10^{-16}$ erg s$^{-1}$ cm$^{-2}$ and observed EW $>393$ {\AA}. We do not detect an emission-line or continuum in the expected wavelength range or anywhere else in the spectrum of SA22-NB816-366911 and SA22-NB816-390412. This may indicate that these sources are variable/transients, as they are also not detected in any of the broad-band images. \cite{Matthee2015} confirmed two of such transients in 0.9 deg$^2$ of similar NB data. Hence, it is not unlikely that our selection picked up three transients in the 3.6 deg$^2$ coverage (see also \citealt{Hibon2010}).

Although we do not detect a clear emission line in the NB816 wavelength coverage in SA22-NB816-508969 and SA22-NB816-360178, we detect a faint trace of continuum in the center of the slits. For SA22-NB816-508969 this continuum is detected at low significance, making it challenging to classify the object. The continuum features, such as the peak-wavelength, of SA22-NB816-360178 resemble those of a star with an effective temperature of $T\approx3500-3700$ K, or a K or M-type star \citep{Kurucz1992}. This interpretation is also strengthened by the point-like morphology in the available imaging. 

The X-SHOOTER spectrum reliably confirms SR6\footnote{SA22 Redshift 6, the brightest LAE at $z=5.7$ in the SA22 field.} as a Lyman-$\alpha$ emitter at $z=5.676\pm0.001$ (using the peak of Ly$\alpha$), due to the asymmetric line-profile (see Fig. $\ref{fig:spectra}$) and non-detection of flux blue-wards of the line. After correcting the VIS spectrum for slit losses of 56 \% (estimated from NB imaging), we measure a line-flux of $7.6\pm0.4\times10^{-17}$ erg s$^{-1}$ cm$^{-2}$, consistent within the errors with the NB estimate of $9.2\pm1.2\times10^{-17}$ erg s$^{-1}$ cm$^{-2}$. We also identify faint [O{\sc ii}] emission from a foreground source at $z=1.322$ offset by 2.4$''$ in the slit. We discuss the detailed properties of SR6 in \S $\ref{sec:goku}$.

\subsection{NB921 targets - candidate LAEs at $\bf z=6.6$}
Out of the six luminous LAE candidates at $z=6.6$ in the SA22 field, we confirm one as a LAE, while we firmly rule out the others at the expected line-fluxes from the NB921 imaging.

Based on its asymmetric line-profile, the source VR7\footnote{Named after Vera Rubin, and chosen to resemble the name of LAE COSMOS Redshift 7 (CR7, \citealt{Matthee2015}), as it was the fifth (V) luminous LAE confirmed at $z\approx6.6$ by the time of discovery.} is confirmed reliably as a LAE at $z=6.532\pm0.001$ (corresponding to the wavelength of peak Ly$\alpha$ emission, see Fig. $\ref{fig:spectra}$). After correcting for an estimated 54 \% of slit losses, we measure a line-flux of $4.9\pm0.5\times10^{-17}$ erg s$^{-1}$ cm$^{-2}$, which agrees well with the NB estimate of $4.8\pm1.2\times10^{-17}$ erg s$^{-1}$ cm$^{-2}$. We present detailed properties of this source in \S $\ref{sec:VR7}$. 

We do not detect an emission-line or a continuum feature in the VIS spectra of SA22-NB921-D10845, SA22-NB921-W210761, SA22-NB921-W219795, SA22-NB921-W6153 or SA22-NB921-W209855, see Table $\ref{tab:targets}$ for a summary. We measure the sensitivity of the spectra as a function of redshift and velocity width of the line. For a line-width of 200 km s$^{-1}$, the 1$\sigma$ limiting flux for wavelengths within the NB921 filter is $\approx4.5 (3.2)\times10^{-18}$ erg s$^{-1}$ cm$^{-2}$ for sources observed with 1 (2) observing blocks (see Table $\ref{tab:targets}$). The sensitivity decreases by a factor $\approx 3$ for a line-width of 600 km s$^{-1}$. However, even with such broad lines, the expected line-fluxes estimated from NB imaging would have been detected at the $>3\sigma$ level. This means that these sources are likely transients (note that \citealt{Matthee2015} estimated that $\sim6$ transients were likely to be found within their sample), and that we can confidently rule out these six sources as Ly$\alpha$ emitters at $z=6.6$. Therefore our results agree very well with the estimates from \cite{Matthee2015} on the fraction of transient interlopers.

\subsection{Updated number densities of the most luminous LAEs at $\bf z\approx6-7$} \label{sec:density}
Based on the spectroscopic follow-up, we provide a robust update on the number densities of luminous LAEs at $z=5.7-6.6$ and compare those with \cite{Santos2016}. At $z=5.7$, the number density of LAEs with L$_{\rm Ly\alpha} = 10^{43.6\pm0.1}$ erg s$^{-1}$ is $10^{-5.26^{+0.21}_{-0.17}}$ Mpc$^{-3}$, which is $\approx 0.25$ dex lower than in \cite{Santos2016}. At $z=6.6$, we find that the number density at L$_{\rm Ly\alpha} = 10^{43.4\pm0.1}$ erg s$^{-1}$ is $10^{-4.89^{+0.22}_{-0.15}}$ Mpc$^{-3}$ and $10^{-5.35^{+0.49}_{-0.22}}$ Mpc$^{-3}$ at L$_{\rm Ly\alpha} = 10^{43.6\pm0.1}$ erg s$^{-1}$. We note that all these number densities are consistent with the previous measurements within 1$\sigma$ errors. The results here support little to no evolution in the bright-end of the Ly$\alpha$ luminosity function between $z=5.7-6.6$, and even little to no evolution at  L$_{\rm Ly\alpha} \approx 10^{43.6}$ erg s$^{-1}$ up to $z=6.9$ \citep{Zheng2017}. After rejecting all candidate LAEs with a luminosity similar to CR7 (for which we measure a total luminosity of $8.5\times10^{43}$ erg s$^{-1}$ after correcting for the transmission curve of the NB921 filter), we constrain the number density of CR7-like sources to one per $\gtrsim 5\times10^{6}$ Mpc$^3$. Catalogues of LAEs at $z=5.7$ and $z=6.6$ will be publicly available with the published version of this paper, see Appendix $\ref{sec:catalogs}$.

\begin{figure}
	\includegraphics[width=8.8cm]{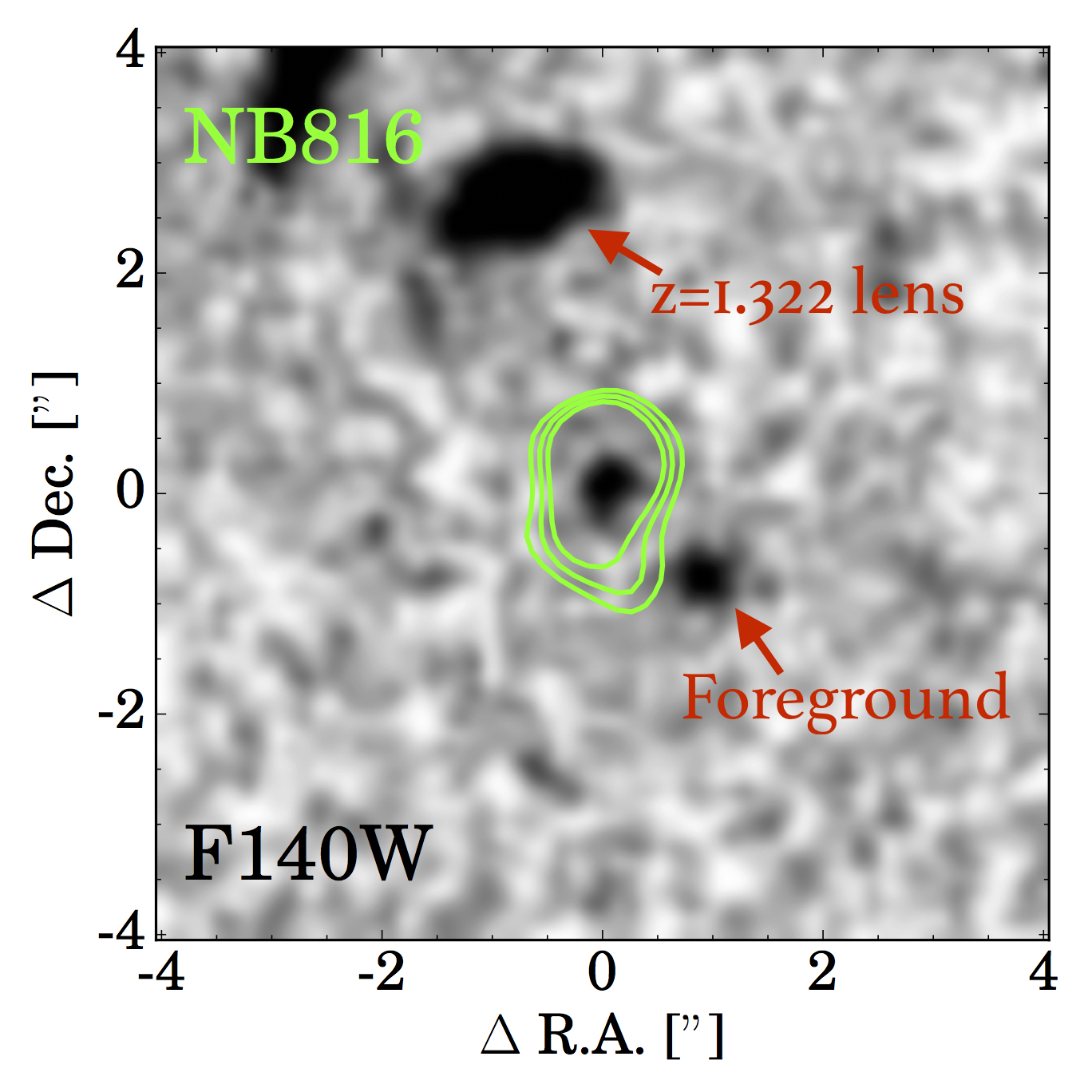} 
    \caption{Rest-frame UV image of SR6 from follow-up with {\it HST} (see \S $\ref{sec:HST_SR6}$). Green contours (at 3, 4 and 5$\sigma$ level) highlight the spatial scales at which we detect Ly$\alpha$ emission in the NB816 filter. The background image is the F140W image, which traces rest-frame wavelengths of $\sim2000$ {\AA}. We note that the PSF of the NB816 imaging is significantly larger than the F098M imaging. SR6 is clearly detected in {\it HST} imaging, resulting in a (magnification corrected) UV luminosity of M$_{1500} = -21.1\pm0.1$ based on the F098M magnitude. {\it HST} imaging also reveals a foreground source at $\sim1$'' that can be identified in the ground-based optical imaging, which could slightly contribute to the flux measured in NB816, explaining why the flux inferred from the NB is slightly higher than the flux inferred from spectroscopy. }
    \label{fig:thumb_goku}
\end{figure}

\section{Properties of newly confirmed LAEs} \label{sec:properties}
\subsection{SR6}\label{sec:goku}
SR6 is robustly confirmed to be a luminous Ly$\alpha$ emitter at $z=5.676\pm0.001$ (Fig. $\ref{fig:spectra}$). We measure a Ly$\alpha$ line-width of $v_{\rm FWHM} = 236\pm16$ km s$^{-1}$ and Ly$\alpha$ luminosity of $2.7\pm0.2\times10^{43}$ erg s$^{-1}$. We do not detect continuum in the X-SHOOTER spectrum (with a 1$\sigma$ depth of $3.0\times10^{-19}$ erg s$^{-1}$ cm$^{-2}$ {\AA}$^{-1}$, smoothed per resolution element), such that we can only provide a lower limit on the EW, which is EW$_0 \gtrsim 250$ {\AA}. Based on \cite{Kashikawa2006}, we quantify the line-asymmetry with the S-statistic and weighted skewness parameters, for which we measure $0.69\pm0.05$ and $9.7\pm0.8$ {\AA}, respectively, similar to other confirmed LAEs.

The foreground [O{\sc ii}] emitter identified in the slit at $z=1.322\pm0.001$, spatially offset by 2.4$''$, is slightly magnifying SR6. We follow \cite{McLure2006} to compute the magnification from galaxy-galaxy lensing as follows:
\begin{equation}
\mu = \frac{d_{\rm proj.}}{d_{\rm proj.}- \theta_E},
\end{equation}
where $\mu$ is the magnification, $d_{\rm proj.}$ is the projected separation in arcsec and $\theta_E$ the Einstein radius in arcsec. Under the assumption of a Singular Isothermal Sphere, we compute $\theta_E$ as follows \citep[e.g.][]{FortMellier1994}:
\begin{equation}
\theta_E = 30'' (\frac{\sigma_{1D}}{1000 \rm \, km \,s^{-1}})^2 \frac{D_{ds}}{D_s},
\end{equation}
where $\sigma_{1D}$ is the one-dimensional velocity dispersion of the foreground source, $D_{ds}$ is the angular diameter distance from foreground source to the background source and $D_s$ the angular diameter distance from observer to the background source. Using the measured $\sigma_{1D} = 130\pm20$ km s$^{-1}$, we estimate $\theta_E = 0.24''$, resulting in a small magnification of $\mu=1.1$. We note that additional magnification by other foreground-sources is possible (for example by a faint source separated by $\sim1''$, see Fig. $\ref{fig:thumb_goku}$), although these sources are likely lower mass due to their faintness, resulting in a further negligible magnification. This results in a magnification corrected Ly$\alpha$ luminosity of $2.5\pm0.3\times10^{43}$ erg s$^{-1}$.

After confirming Ly$\alpha$, we investigate the optical and near-infrared spectra for the presence of other emission-lines in the rest-frame UV. In particular, we search for N{\sc v}, C{\sc iv}, He{\sc ii}, O{\sc iii}] and C{\sc iii}]\footnote{In vacuum, the wavelengths of these lines are N{\sc v}$_{\lambda\lambda} =1239,1243$ {\AA}, C{\sc iv}$_{\lambda\lambda} =1548,1551$ {\AA}, He{\sc ii}$_{\lambda} =1640$, O{\sc iii}]$_{\lambda\lambda} =1661,1666$ {\AA} and C{\sc iii}]$_{\lambda\lambda} =1907,1909$ {\AA}.}, and check for any other significantly detected potential line -- but we do not detect any above 3$\sigma$ significance. We measure limiting line-fluxes at the positions of the expected lines for a range of line-widths. For a line-width of $\sim100-250$ km s$^{-1}$ and typical velocity offset with respect to Ly$\alpha$ of $-200$ km s$^{-1}$, we find a 2$\sigma$ limit of $2.0\times10^{-17}$ erg s$^{-1}$ cm$^{-2}$ for N{\sc v} after correcting for the same slit losses as Ly$\alpha$ (corresponding to EW$_0 < 48$ {\AA}).
For the other lines (observed in the NIR slit), we estimate slit losses of 59 \%. This assumes that these lines are emitted over the same spatial scales as Ly$\alpha$. Because sources are un-detected in the NIR continuum, we can not estimate slit losses from the continuum emission itself. As Ly$\alpha$ is likely emitted over a larger spatial scale \citep[e.g.][]{Wisotzki2015}, slit losses for the other rest-UV lines may be over-estimated (except potentially for C{\sc iv} which is also a resonant line). Our upper limits are on the conservative side if this is indeed the case. For similar widths and offsets as N{\sc v}, we measure 2$\sigma$ limiting line-fluxes of (7.3, 3.6, 3.5, 3.9) $\times10^{-17}$ erg s$^{-1}$ cm$^{-2}$ for (C{\sc iv}, He{\sc ii}, O{\sc iii}], C{\sc iii}]), corresponding to EW$_0$ $<(174, 86, 84, 93)$ {\AA}, respectively. These limits are not particularly strong because all lines are either observed around strong sky OH lines or at low atmospheric transmission, but also due to our modest exposure time and conservative way of measuring noise.  

\subsubsection{HST follow-up} \label{sec:HST_SR6}
We observed SR6 with our ongoing {\it HST}/WFC3 follow-up program (PI Sobral, program 14699), and is detected in the F098M and F140W filter, see e.g. Fig. $\ref{fig:thumb_goku}$, with a total integration time of 4076\,s and 3176\,s. The source is marginally resolved, consists of a single component that is separated by $\approx0.2''$ from the peak Ly$\alpha$ flux. We measure magnitudes of F098M=$25.68\pm0.13$ and F140W=$25.60\pm0.10$ in a 0.4$''$ aperture. Correcting for magnification, this results in M$_{1500} = -21.1\pm0.1$, which corresponds to a dust-uncorrected SFR $\approx10$ M$_{\odot}$ yr$^{-1}$ and is thus a $M_{UV}^{\star}$ source at that redshift \citep{Bouwens2015}. Following the calibration from \cite{Schaerer2015}, we estimate a stellar mass of M$_{\rm star} \approx 4\times10^9$ M$_{\odot}$. The galaxy has a moderately blue UV slope, $\beta=-1.78\pm0.45$. In both {\it HST} filters, we measure a size of r$_{1/2} = 0.8\pm0.2$ kpc using SExtractor (corrected for PSF broadening following e.g. \citealt{CurtisLake2016,Ribeiro2016}).  We use the {\it HST} photometry to estimate the continuum around Ly$\alpha$ and measure Ly$\alpha$ EW$_0 = 802\pm155$ {\AA}. While the SFR, size and UV slope are typical, and not very different from UV selected galaxies at $z\approx6-7$ \citep[e.g.][]{Bowler2017}, the extremely high Ly$\alpha$ EW is challenging to explain with simple stellar populations \citep[e.g.][]{CharlotFall1993}, indicating an elevated production rate of ionising photons. Such high EWs are also found in numerous other Ly$\alpha$ surveys \citep[e.g.][]{MalhotraRhoads2002,Hashimoto2016}, although we note that those sources are typically of fainter luminosity. High EWs may be explained by extremely low metallicity stellar populations with young ages \citep[e.g.][]{Schaerer2003}. Other explanations include AGN activity and contributions from cooling radiation \citep{Rosdahl2012} and shocks \citep{Taniguchi2015}. However, these processes typically result in more extended Ly$\alpha$ emission, which is not observed with the current observational limits. Ly$\alpha$ EW may also be boosted in a clumpy ISM \citep[e.g.][]{Duval2014,Gronke2014}, but we note that measurements of the UV slope indicate little dust.

\begin{figure}
	\includegraphics[width=8.6cm]{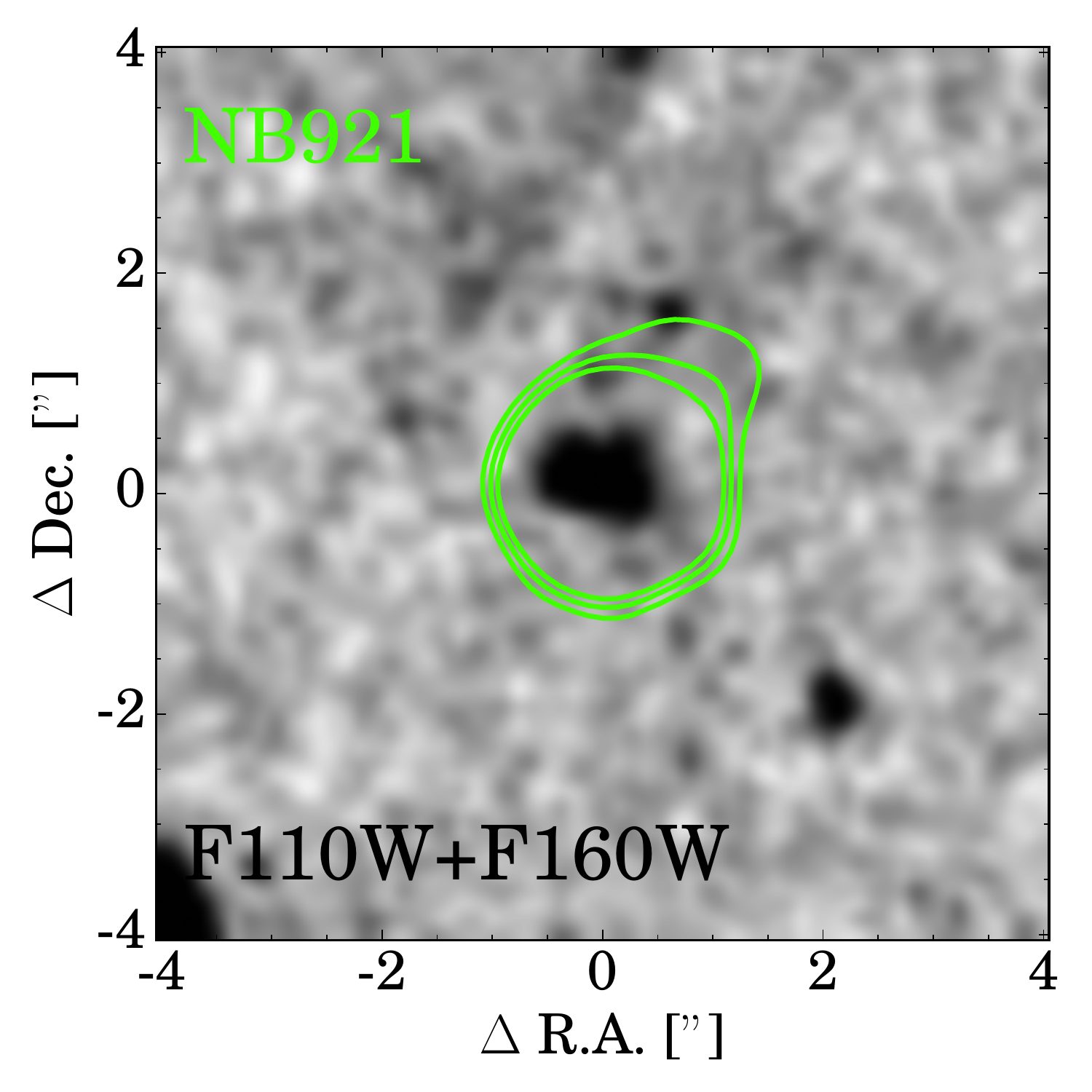} 
    \caption{Rest-frame UV (F110W+F160W) image of VR7, which traces rest-frame wavelengths $\sim 1500$ {\AA}. The green contours show the Ly$\alpha$ emission measured from NB921 (at 3, 4 and 5$\sigma$ level).  VR7 is elongated in the UV continuum, possibly due to two merging components. The absolute UV magnitude measured within a 2$''$ aperture centred on the Ly$\alpha$ peak is M$_{1500}=-22.5\pm0.1$, see $\ref{sec:HST_VR7}$.} 
    \label{fig:thumb_VR7}
\end{figure}

\subsection{VR7}\label{sec:VR7}
The source VR7 is a Ly$\alpha$ emitter at $z=6.532\pm0.001$, see Fig. $\ref{fig:spectra}$, with a Ly$\alpha$ luminosity of $2.4\pm0.2\times10^{43}$ erg s$^{-1}$. We do not detect continuum, allowing us to place a lower limit on the equivalent width of EW$_0 > 196$ {\AA}. The Ly$\alpha$ line-width is $v_{\rm FWHM} = 340\pm14$ km s$^{-1}$, the S-statistic is $0.33\pm0.04$, resulting in a Skewness of $6.9\pm0.8$ {\AA}. This skewness is similar to those measured in fainter LAEs at $z=6.5$ by \cite{Kashikawa2011}.

We do not detect any emission-line besides Ly$\alpha$ in the optical or near-infrared spectrum, and place the following 2$\sigma$ limits (assuming line-widths of $\sim200$ km s$^{-1}$, narrower than Ly$\alpha$, but similar to other studies): (1.0, 2.3, 2.3, 2.2, 2.1) $\times10^{-17}$ erg s$^{-1}$ cm$^{-2}$ for (N{\sc v}, C{\sc iv}, He{\sc ii}, O{\sc iii}], C{\sc iii}]), see Table $\ref{tab:properties}$. These error estimates are also measured including sky OH lines, even though the lines themselves may avoid skylines, and are thus conservative. Assuming a continuum level of $1.3\times10^{-19}$ erg s$^{-1}$ cm$^{-2}$ {\AA}$^{-1}$, these flux limits translate into EW$_0$ limits of $<$(9, 21, 21, 20, 19) {\AA}, respectively.

\begin{table}
\centering
\caption{Measurements of SR6 and VR7. Luminosity and EW are measured through spectroscopy. SFR$_{\rm UV}$ is based on the absolute UV magnitude, assuming negligible dust attenuation and a Chabrier IMF. Stellar mass is based on UV luminosity, following a calibration based on SED models presented in \citet{Schaerer2015}. $\xi_{ion}$ is computed as described in \S $\ref{sec:xion}$, which assumes f$_{\rm esc, Ly\alpha} = 100$ \% for both sources, and is thus a lower limit. Line-flux 2$\sigma$ limits are in 10$^{-17}$ erg s$^{-1}$ cm$^{-2}$ and EW$_0$ limits are in {\AA}. } 
\begin{tabular}{lrr}
\hline
Measurement &SR6 & VR7  \\  \hline
$z_{spec, \rm Ly\alpha}$ & $5.676\pm0.001$  & $6.532\pm0.001$ \\
L$_{\rm Ly\alpha}$/10$^{43}$ erg s$^{-1}$ & $2.5\pm0.3$  & $2.4\pm0.2$ \\
EW$_{0, \rm spec}$/{\AA} & $>250$ {\AA} & $>196$ {\AA} \\
 EW$_{0, \rm spec+phot}$/{\AA}  & $802\pm155$ {\AA} & $207\pm10$ {\AA} \\
v$_{\rm FWHM, Ly\alpha}$/km s$^{-1}$ & $236\pm16$ & $340\pm14$\\
Skewness/{\AA} & $9.7\pm0.8$  & $6.9\pm0.8$ \\
M$_{1500}$ & $-21.1\pm0.1$ & $-22.5\pm0.1$ \\
SFR$_{\rm UV}$/M$_{\odot}$ yr$^{-1}$ & 10 & 38 \\
M$_{\rm star}$/M$_{\odot}$ & $4\times10^9$ & $1.7\times10^{10}$  \\
log$_{10} (\xi_{ion}$/Hz erg$^{-1}$) & $\gtrsim25.25\pm0.23$ & $\gtrsim24.66\pm0.17$  \\  
$\beta$ & $-1.78\pm0.45$ & $-1.97\pm0.31$ \\
r$_{1/2}$/kpc & $0.9\pm0.1$ & $1.7\pm0.1$ \\
f$_{\rm NV}$ (EW$_{0, \rm NV}$) & $<2.0$ $(<48)$ & $<1.0$ $(<9)$ \\
f$_{\rm CIV}$ (EW$_{0, \rm CIV}$) & $<7.3$ $(<174)$ & $<2.3$ $(<21)$ \\
f$_{\rm HeII}$ (EW$_{0, \rm HeII}$) & $<3.6$ $(<86)$ & $<2.3$ $(<21)$ \\
f$_{\rm OIII]}$ (EW$_{0, \rm OIII]}$) & $<3.5$ $(<84)$ & $<2.2$ $(<20)$ \\
f$_{\rm CIII]}$ (EW$_{0, \rm CIII]}$) & $<3.9$ $(<93)$ & $<2.1$ $(<19)$ \\

\hline
\end{tabular}
\label{tab:properties}
\end{table}

\subsubsection{HST follow-up} \label{sec:HST_VR7}
VR7 is detected at $\approx3\sigma$ significance in the UKIDSS DXS $J$ band imaging ($J=24.2$), resulting in an absolute UV magnitude of M$_{1500} = -22.5\pm0.2$. This luminosity places the source in the transition region between luminous galaxies and faint AGN \citep[e.g.][]{Willott2009,Matsuoka2016} and is $\approx0.3$ dex brighter than CR7 \citep[e.g.][]{Sobral2015}. We also obtained {\it HST}/WFC3 imaging in the F110W and F160W filters (PI Sobral, program 14699), with integration times of 2612\,s and 5223\,s. These observations reveal a relatively large elongated galaxy (r$_{1/2} = 1.7\pm0.1$ kpc, elongation of 1.4), with F110W=$24.33\pm0.09$ and F160W=$24.32\pm0.10$ in a 0.6$''$ aperture, see Fig. $\ref{fig:thumb_VR7}$. We constrain the UV slope to $\beta=-1.97\pm0.31$. The UV luminosity corresponds to a SFR of 38 M$_{\odot}$ yr$^{-1}$, under the assumptions that the UV luminosity originates from star-formation (as noted above, we do not detect any signs of AGN activity such as C{\sc iv} or Mg{\sc ii} emission at the current detection limits), a Chabrier IMF and that dust attenuation is negligible. Based on the calibration from \cite{Schaerer2015}, the stellar mass is  M$_{\rm star} \approx 1.7\times10^{10}$ M$_{\odot}$. Similar to SR6, we constrain the Ly$\alpha$ EW using {\it HST} photometry, and find EW$_0 = 207\pm10$ {\AA}, which is higher than the typically assumed maximum EW possible due to star-formation \citep[e.g.][]{CharlotFall1993}, and indicates strongly ionising properties. Because of these properties, VR7 is an ideal target for further detailed follow-up observations. 

\begin{table}
\centering
\caption{Compilation of Ly$\alpha$ line-widths of spectroscopically confirmed LAEs at $5.6<z<6.6$ included in Fig. $\ref{fig:vfwhm-lya}$. These sources are included in Fig. $\ref{fig:vfwhm-lya}$ in addition to the spectroscopically confirmed LAEs from \citet{Hu2010}, \citet{Ouchi2010}, \citet{Kashikawa2011} and \citet{Shibuya2017}. More detailed information on these sources is included in Table $\ref{tab:compilation}$. }
\begin{tabular}{lrrrrrrrrp{3cm}}
\hline
ID & Redshift & v$_{\rm FWHM, Ly\alpha}$  \\ 
 & & km s$^{-1}$ \\ \hline
SGP 8884 & 5.65 & $250\pm30$ \\ 
SR6 & 5.67 & $236\pm16$ \\
Ding-1 &  5.70 & $340\pm100$ \\   
S11 5236 &  5.72 & $300\pm30$ \\
VR7 & 6.53 & $340\pm14$ \\
MASOSA & 6.54 & $386\pm30$ \\ 
Himiko  & 6.59 & $251\pm21$ \\ 
COLA1 & 6.59 & $194\pm42$ \\
CR7 & 6.60 & $266\pm15$ \\ \hline

\end{tabular}
\label{tab:linewidths}
\end{table}

\section{Discussion} \label{sec:discussion}
\subsection{The evolution of Ly$\alpha$ line-widths} \label{sec:widths}
In order to investigate the nature of luminous LAEs at $z=5.7-6.6$ using their Ly$\alpha$ line-profile, we compare the measurements with a reference sample of luminous LAEs at $z\approx2-3$ (Sobral et al. in prep). These comparison sources have been selected with wide-area narrow-band surveys \citep[e.g.][]{Sobral2016,Matthee2017Bootes}, and we match the minimum EW$_0$ criterion to $>20$ {\AA}. Even when we exclude broad-line AGN from the $z\approx2-3$ sample, we find that luminous LAEs at $z=5.7-6.6$ have Ly$\alpha$ line-widths (typically $290\pm20$ km s$^{-1}$) that are a factor 2-3 narrower than those at $z\approx2-3$. These sources at lower redshift are a mix of narrow-line AGN and star-forming galaxies. This indicates that, besides non-detections of AGN associated lines as C{\sc iv} or Mg{\sc ii}, the Ly$\alpha$ lines do not clearly indicate AGN activity in luminous LAEs at $z=5.7-6.6$.

Due to resonant scattering, the presence of neutral hydrogen broadens Ly$\alpha$ emission lines \citep[e.g.][]{Kashikawa2006,Dijkstra2014}. Theoretically, \cite{HaimanCen2005} show that the observed Ly$\alpha$ line FWHM increase mostly at faint luminosities, L$_{\rm Ly\alpha} \approx10^{42}$ erg s$^{-1}$, with a more prominent evolution with higher neutral fraction and narrower intrinsic line-width. Therefore, evolution in the observed Ly$\alpha$ profiles at $z\gtrsim6$ may provide hints on how reionisation happened. We investigate whether we find evidence for increasingly broad Ly$\alpha$ profiles as a function of redshift, by controlling for differences in Ly$\alpha$ luminosities.

\begin{figure*}
	\includegraphics[width=14cm]{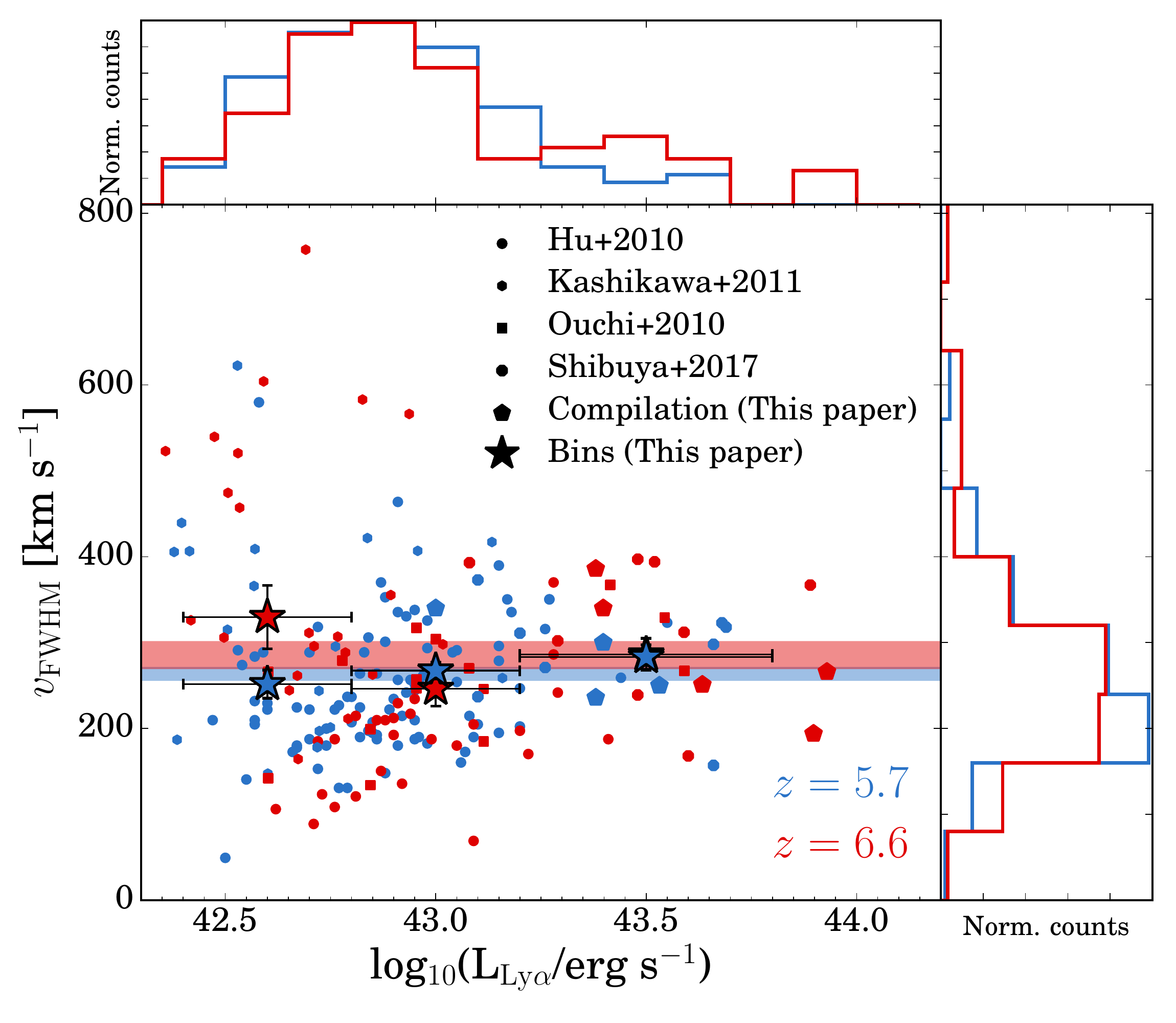}
    \caption{Ly$\alpha$ line-widths as a function of Ly$\alpha$ luminosity. Blue points show LAEs at $z=5.7$, while red points show LAEs at $z=6.6$. The red and blue horizontal bands indicate the mean line-widths and its error, while stars show the mean in bins of Ly$\alpha$ luminosity. At $z=5.7$ Ly$\alpha$ line-widths increase slightly with increasing luminosity. While the average over the full sample indicates no significant evolution in line-widths from $z=5.7-6.6$, the binned-averages indicate that line-widths of faint LAEs at $z=6.6$ are a factor $\sim1.2$ higher than at $z=5.7$.  }
    \label{fig:vfwhm-lya}
\end{figure*}

In Fig. $\ref{fig:vfwhm-lya}$ we show the dependence of Ly$\alpha$ line-width on Ly$\alpha$ luminosity for samples at $z=5.7$ and $z=6.6$. We include samples from \cite{Hu2010}, \cite{Ouchi2010}, \cite{Kashikawa2011}, \cite{Shibuya2017} and the compilation of Ly$\alpha$ selected sources from Table $\ref{tab:linewidths}$, which includes the two sources confirmed in this paper. This compilation also includes the luminous LAEs at $z=5.7$ discovered by \cite{Westra2006}, studied in detail in \cite{Lidman2012}, and the double-peaked LAE COLA1 at $z=6.593$ discovered by \cite{Hu2016}\footnote{In our analysis of COLA1, we find that it is detected at $>5\sigma$ in the public HST F814W imaging \citep{Koekemoer2007}, with a magnitude of $26.2\pm0.2$. Even though the F814W filter has significant transmission above 920 nm, this magnitude indicates that a fraction of the flux density measured in F814W originates from $\lambda_0<1216$ {\AA}. The F814W imaging also shows a neighbouring source within the PSF-FWHM of the NB921 imaging (data from Subaru program S13A-057; \citealt{Sobral2013}), indicating that the Ly$\alpha$ luminosity may be over-estimated. There are also $\sim2 \sigma$ detections in the $B$ and $V$ band Suprime-Cam images. These detections are unexpected for a source at $z=6.6$ (because they trace below the Lyman break), and could indicate that the emission-line is the [O{\sc ii}]$_{3727,3729}$ doublet at $z=1.477$ (similar to the photometric redshift of the source in the \citealt{Laigle2016} catalogue). On the other hand, while the double-peak separation in the spectrum presented in \cite{Hu2016} may be explained with the [O{\sc ii}] doublet, the asymmetric red wing challenges this explanation. Thus, currently none of the scenarios is completely satisfactory. Follow-up observations in the NIR are required to fully distinguish between these scenarios.}.
While there is significant scatter, there are some interesting results. Firstly, the binned results indicate that line-widths at $z=5.7$ increase slightly with increasing Ly$\alpha$ luminosity (at $\approx3\sigma$ significance, see also \citealt{Hu2010}), while this is not necessarily the case at $z=6.6$. In order to estimate the error on the bins as conservatively as possible, we combine the formal error ($\sigma$/$\sqrt N$, where $\sigma$ is the observed standard deviation and $N$ the number of sources in each bin) and the 1$\sigma$ uncertainty on the mean estimated through bootstrap resampling the sample in the bins 1000 times in quadrature. By fitting a linear relation through the binned points at $z=5.7$, we find that line-width increases with luminosity as:
\begin{equation}
v_{\rm FWHM} \rm  = 35^{+16}_{-13} \, log_{10}(\frac{L_{\rm Ly\alpha}}{10^{43} \rm erg s^{-1}}) +267^{+11}_{-11}\, km\, s^{-1}. 
\end{equation}

Secondly, the average values in bins of Ly$\alpha$ luminosity indicate that LAEs with luminosities L$_{\rm Ly\alpha} = 10^{42.4-42.8}$ erg s$^{-1}$ have broader line-widths at $z=6.6$ ($v_{\rm FWHM} \approx 330$ km s$^{-1}$) than at $z=5.7$ ($v_{\rm FWHM} \approx 250$ km s$^{-1}$). We test the significance of these results by taking the uncertainties due to the limited sample size into account as follows. We bootstrap the sample in the low luminosity bins at both $z=5.7$ and $z=6.6$ 1000 times and we compute the mean $v_{\rm FWHM}$ in each realisation. The 1$\sigma$ error on the mean is then the standard deviation of these 1000 measurements. At $z=5.7$ we find $v_{\rm FWHM} = 252 \pm 17$ (error on mean) $\pm112$ (dispersion) km s$^{-1}$, while at $z=6.6$ we find $v_{\rm FWHM} = 323 \pm 36$ (error on mean) $\pm192$ (dispersion) km s$^{-1}$. This means that the offset is only marginally significant. We also perform a Kolmogorov-Smirnov test on 1000 realisations of the sample where we have perturbed each measured $v_{\rm FWHM}$ with its uncertainty assuming that the uncertainty is gaussian. We find a mean P-value of $0.12\pm0.05$ and a KS-statistic of $0.27\pm0.02$. This means that the two distributions are not drawn from the same parent distribution at $\approx85$ \% confidence level. This difference in the line-widths between the samples at $z=5.7$ and $z=6.6$ resembles the prediction from \cite{HaimanCen2005} and may be used to constrain the neutral fraction of the IGM. As the dispersion is relatively large and the difference is significant at only $\approx 85$ \% confidence level, larger samples are required to better constrain this evolution.

The trends that Ly$\alpha$ line-width increases slightly with luminosity at $z=5.7$ and that the faintest LAEs may have broader Ly$\alpha$ lines at higher redshift, may explain why neither \cite{Hu2010}, \cite{Ouchi2010} or \cite{Kashikawa2011} report increasing Ly$\alpha$ line-widths between $z=5.7-6.6$. This is because they only studied the average over all luminosities (which does not change significantly), or probed a different Ly$\alpha$ luminosity regime. Interestingly, the luminosity at which line-widths may increase might correspond to the luminosity where the number density (at fixed Ly$\alpha$ spatial scale) drops most strongly between $z=5.7-6.6$ \citep{Matthee2015}, and where there is relatively more extended Ly$\alpha$ emission at $z=6.6$ than at $z=5.7$ \citep{Santos2016}. This strengthens the idea that we are witnessing the effect of patchy reionisation affecting the number densities, line-widths and spatial extents of faint Ly$\alpha$ emitters at $z=6.6$.

\begin{table*}
\centering
\caption{Compilation of Ly$\alpha$ luminosities, EWs, absolute UV magnitudes and line-ratios between Ly$\alpha$ and rest-frame UV lines. Galaxies are either categorised as Ly$\alpha$ (narrow-band) selected, or UV (Lyman-break) selected, and are ordered by increasing redshift (see Table $\ref{tab:references}$). For the doublets C{\sc iv} C{\sc iii}] and O{\sc iii}], we use the combined flux. Upper limits are at the 2$\sigma$ level.}
\begin{tabular}{lrrrrrrrr}
\hline
ID & L$_{\rm Ly\alpha}$ & EW$_{0, \rm Ly\alpha}$ & M$_{1500}$ & N{\sc v}/Ly$\alpha$ & C{\sc iv}/Ly$\alpha$ & He{\sc ii}/Ly$\alpha$ & O{\sc iii}]/Ly$\alpha$ & C{\sc iii}]/Ly$\alpha$   \\ 
& erg s$^{-1}$ & {\AA} & AB & & & & &  \\ \hline 
\bf Ly$\alpha$ selected & &  &  & & &  & &\\
SGP 8884 & $3.4\times10^{43}$ & 166 & - & $<0.01$ & - & - & $<0.09$ & $<0.13$ \\ 
SR6 & $2.5\times10^{43}$ & $>250$ & -21.1 & $<0.26$ & $<0.96$ & $<0.47$ & $<0.46$ & $<0.51$ \\
Ding-3 &  $0.7\times10^{43}$ &62 & -20.9 & - & - & - & - & $<0.11$  \\  
Ding-4 &  $0.2\times10^{43}$ &106 & -20.5 & - & - & - & - & $<0.31$  \\  
Ding-5 &  $2\times10^{43}$ &79 & -20.5 & - & - & - & - & $<0.05$ \\  
Ding-2 &  $0.2\times10^{43}$ & - & -22.2 & - & - & - & - & $<0.31$  \\  
Ding-1 &  $1\times10^{43}$ & 21 & -22.2 & - & - & - & - & 0.09  \\  
J233408 &  $4.8\times10^{43}$ & $>260$ & $>-20.8$ & $<0.05$ & $0.08$ & $<0.01$ & $<0.01$ &- \\	

S11 5236 & $2.5\times10^{43}$ & 160 & - & $<0.03$ & $<0.13$ & - & $<0.21$ & $<0.18$  \\
J233454 & $4.9\times10^{43}$ & 217 & -21.0 &  $<0.05$ & $<0.01$ & $<0.01$ & $<0.01$ &- \\
J021835 &  $4.6\times10^{43}$ & 107 & -21.7&  $<0.07$ & $<0.02$ & $<0.03$ & $<0.01$ &- \\
WISP302 & $4.7\times10^{43}$ & 798 & -19.6 & - & - & $<0.41$ & - & $<0.29$  \\  
VR7 & $2.4\times10^{43}$ & $>196$ & -22.5 & $<0.16$ & $<0.36$ & $<0.35$ & $<0.35$ & $<0.33$ \\
LAE SDF-LEW-1 & $1\times10^{43}$ & 872 & $>-22$ & - & $<0.01$ & $<0.02$ & - & - \\ 
J162126 &  $7.8\times10^{43}$ & 99 & -20.5 &  $<0.05$ & $<0.01$ & $<0.02$ & $<0.01$ &- \\
J160940 &  $1.9\times10^{43}$ & $>31$ & $>-22.1$ &  $<0.14$ & $<0.19$ & $<0.30$ & $<0.49$ &- \\
J100550 &  $3.9\times10^{43}$ & $>107$ & $>-21.5$ &  $<0.08$ & $<0.01$ & $<0.01$ & $<0.03$ &-\\
J160234 &  $3.3\times10^{43}$ & 81 & -21.9&  $<0.11$ & $<0.12$ & $<0.16$ & $<0.23$ &- \\
Himiko  & $4.3\times10^{43}$ & 65 & -22.1& $<0.03$ & $<0.10$ & $<0.05$ & - & $<0.08$ \\ 
CR7 (recalibrated) & $8.5\times10^{43}$& 211 & -22.2 & $<0.03$ & $<0.12$ & $0.14\pm0.06$ & $<0.09$ & $<0.11$  \\ 
 & &  &  & & &  & &\\
\bf UV selected & &  &  & & & & & \\
A383-5.2  & $0.7\times10^{43}$ & 138 & -19.3 & - & - & - & - & $0.05\pm0.01$  \\ 
RXCJ2248.7-4431-ID3 & $0.3\times10^{43}$ & 40 & -20.1 & $<0.05$ & $0.42\pm0.12$ & $<0.05$ & $0.13\pm0.04$ & $<0.11$  \\ 
RXCJ2248.7-4431 &  $0.8\times10^{43}$ & 68 & -20.2 & $<0.48$ & $0.45\pm0.12$ & $<0.28$ & $0.31\pm0.12$ & $<0.09$ \\ 
SDF-46975& $1.5\times10^{43}$ & 43 & -21.5 & $<0.13$ & - & - &- & - \\ 
IOK-1 & $1.1\times10^{43}$ & 42 & -21.3 & $<0.17$ & - & $<0.12$ & - & -  \\ 
BDF-521 & $1.0\times10^{43}$ & 64 & -20.6 & $<0.26$ & -  & $<0.16$ & - & - \\ 
A1703\_zd6  & $0.3\times10^{43}$ &65& -19.3 & - & $0.28\pm0.03$ & $<0.07$ & $0.06\pm0.03$* &  - \\  
BDF-3299 & $0.7\times10^{43}$ &50 & -20.6 & $<0.26$ & - &- &- & - \\ 
GLASS-stack & $1\times10^{43}$ & 210* & -19.7 & $<0.4$ & $<0.3$ & $<0.2$ & $<0.2$ & $<0.2$ \\ 
EGS-zs8-2  & $0.5\times10^{43}$ & 9 & -21.9 & - & - & - & - & $<0.41$ \\  
FIGS\_GN1\_1292 & $0.7\times10^{43}$ & 49 & -21.2 & $0.85\pm0.25$ & - & - & - & - \\ 
GN-108036 & $1.5\times10^{43}$ & 33 & -21.8 & $<0.33$ & - & - & - & $0.09\pm0.05$  \\  
EGS-zs8-1 & $1.2\times10^{43}$ & 21 & -22.1 & - & - & - & - & $0.46\pm0.10$ \\  

\hline\end{tabular}
\label{tab:compilation}
\end{table*}

\begin{figure*}
\begin{tabular}{cc}
	\includegraphics[width=8.6cm]{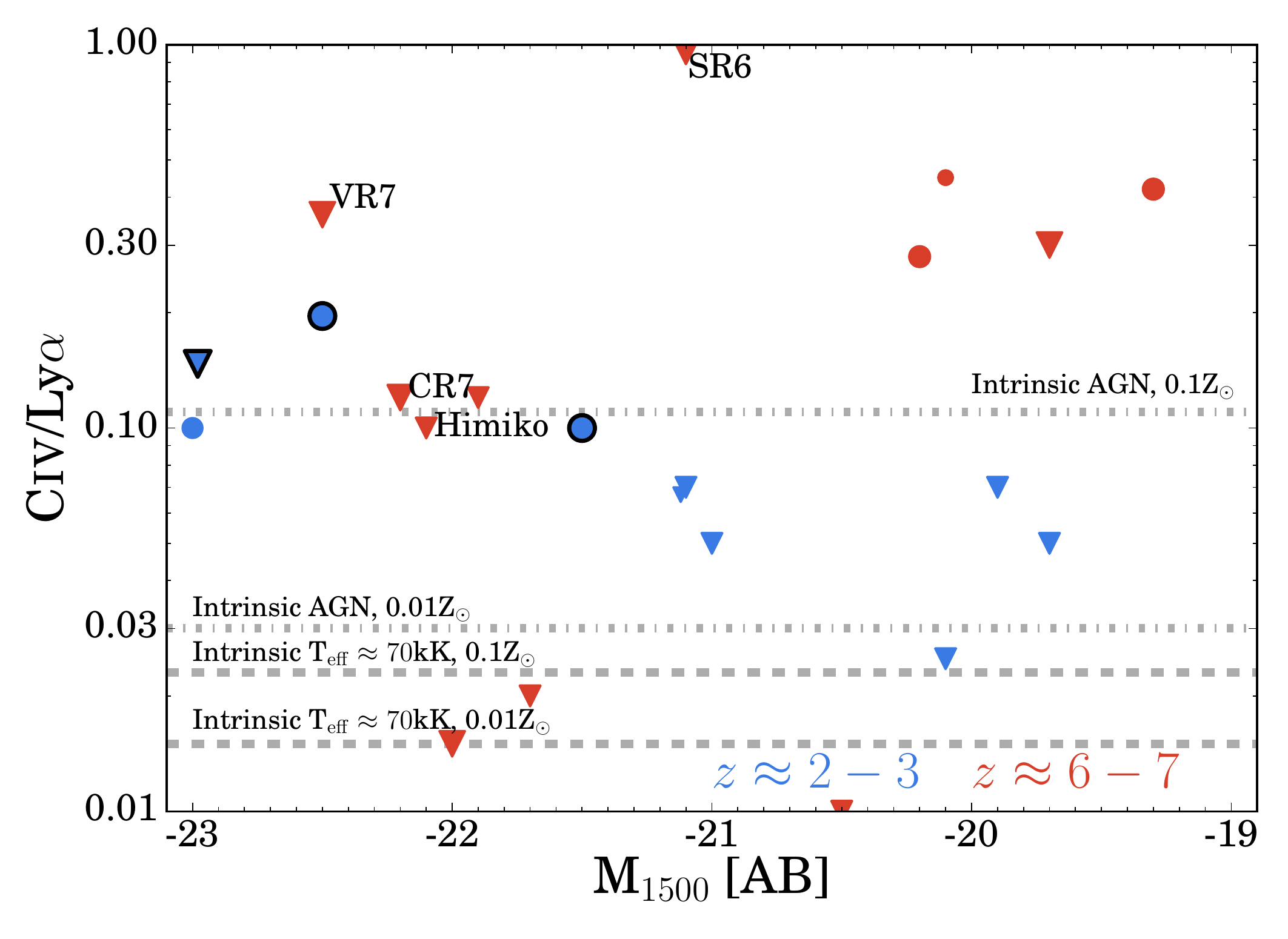}  &
\includegraphics[width=8.6cm]{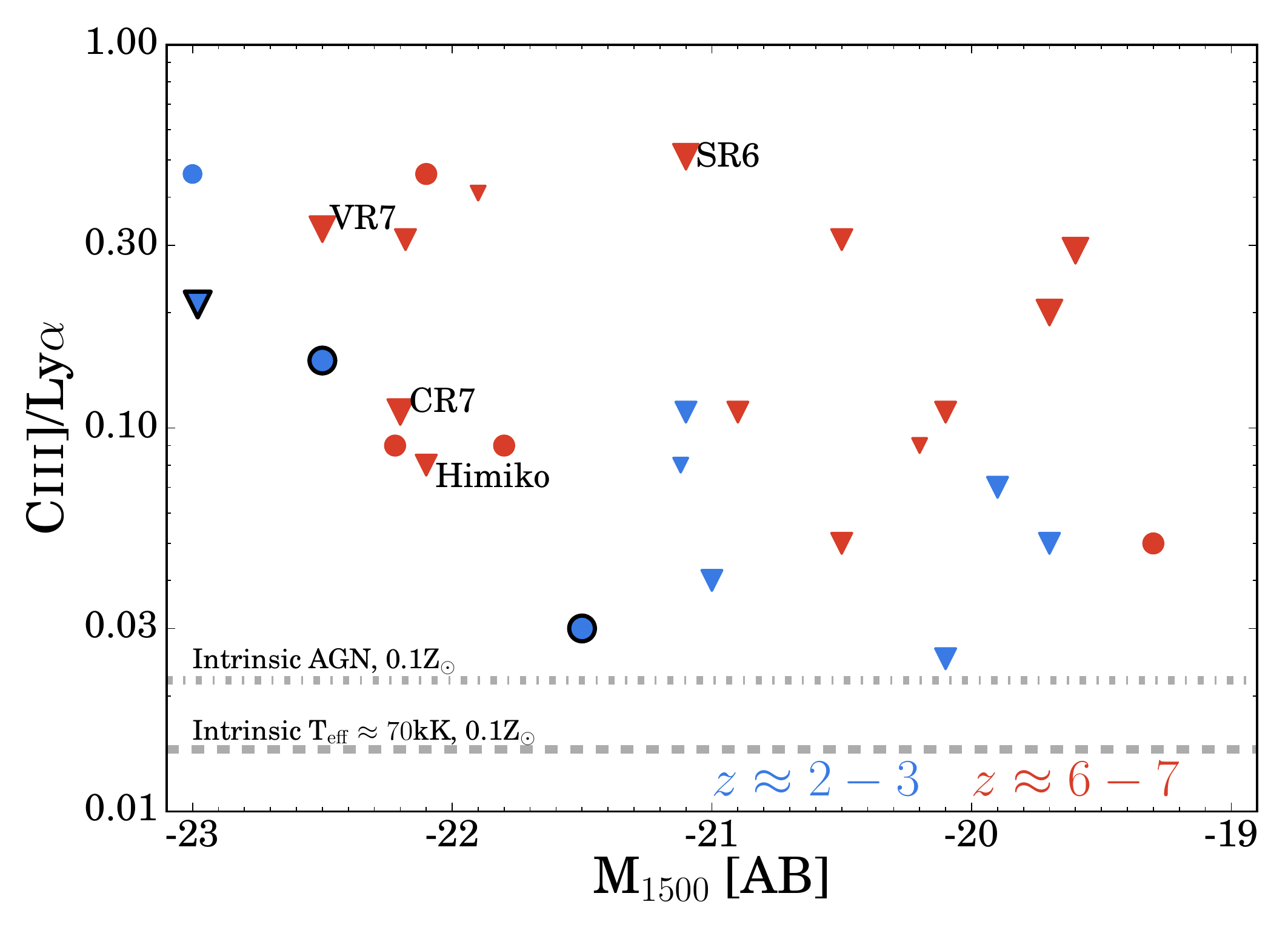}
\end{tabular}
    \caption{Observed C{\sc iv}/Ly$\alpha$ and C{\sc iii}]/Ly$\alpha$ ratios as a function of M$_{1500}$ for luminous LAEs at $z\approx2-3$ (Sobral et al. in prep) and the compilation of LAEs and LBGs at $z\approx6-7$ from Table $\ref{tab:compilation}$ with C{\sc iv} and C{\sc iii}] detections and/or upper limits. Upper limits are shown with down-ward pointing triangles, while detections are shown with circles. We highlight AGN in the $z\approx2-3$ sample with black edges. The symbol sizes increase with increasing Ly$\alpha$ EW. Horizontal lines indicate estimated intrinsic line-ratios (Alegre et al. in prep), assuming a 100 \% Ly$\alpha$ escape fraction. Galaxies with C{\sc iv} detections at $z\approx6-7$ have higher C{\sc iv}/Ly$\alpha$ ratios than LAEs at $z\approx2-3$ with similar UV luminosities. The limits on CR7 and Himiko are comparable to detections of similar sources at $z\approx2-3$, but for which an AGN nature is confirmed. }
    \label{fig:C-ratios}
\end{figure*}

\subsection{UV (metal) line-ratios to Ly$\alpha$} \label{sec:compilation}
As described in \S $\ref{sec:properties}$, no rest-UV metal-lines are detected in SR6 or VR7. Such lines have also not been detected in Himiko \citep{Zabl2015} or CR7 \citep{Sobral2015}. In this section we explore whether this is due to the limited depth of the observations or may be attributed to any peculiar physical condition (for example due to a low metallicity). As a comparison sample, we made a compilation of UV and Ly$\alpha$ selected galaxies at $z\gtrsim6$ for which limits on other UV emission-lines besides Ly$\alpha$ are published, see Table $\ref{tab:compilation}$. These sources have all been spectroscopically confirmed through their Ly$\alpha$ emission. All upper limits are converted to 2$\sigma$ and we compute Ly$\alpha$ luminosities and absolute UV magnitudes based on the published observed magnitudes and fluxes in the case luminosities and absolute magnitudes have not been provided. We show limits on the strength of N{\sc v}, C{\sc iv}, He{\sc ii}, O{\sc iii}] and C{\sc iii}] compared to Ly$\alpha$. A more detailed description on the compiled sample is provided in Appendix $\ref{sec:compilation_info}$. In addition, we also compare our sources with a sample of luminous LAEs at $z\approx2-3$ (Sobral et al. in prep).

\begin{figure*}
\begin{tabular}{cc}
	\includegraphics[width=8.6cm]{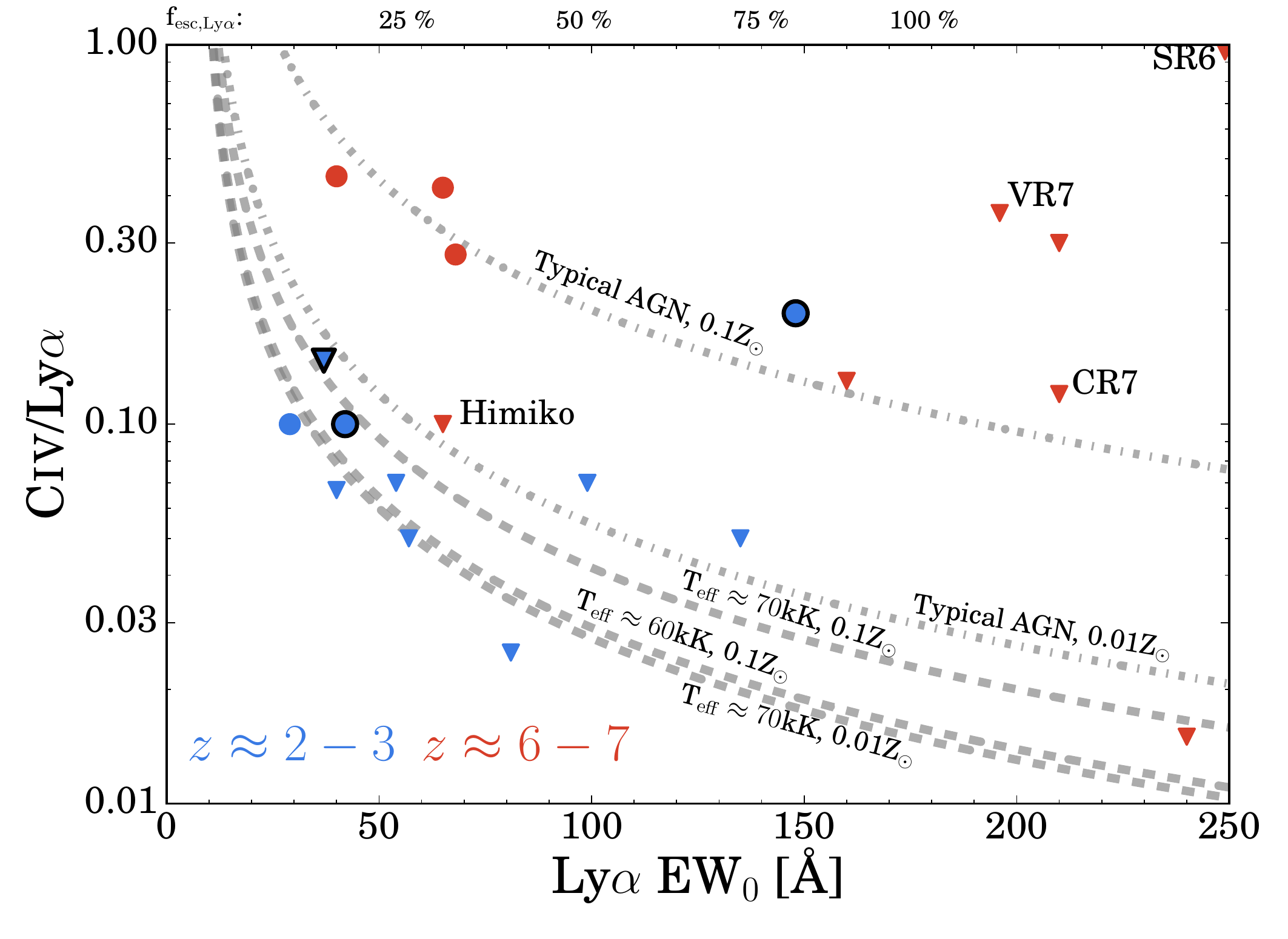}  &
\includegraphics[width=8.6cm]{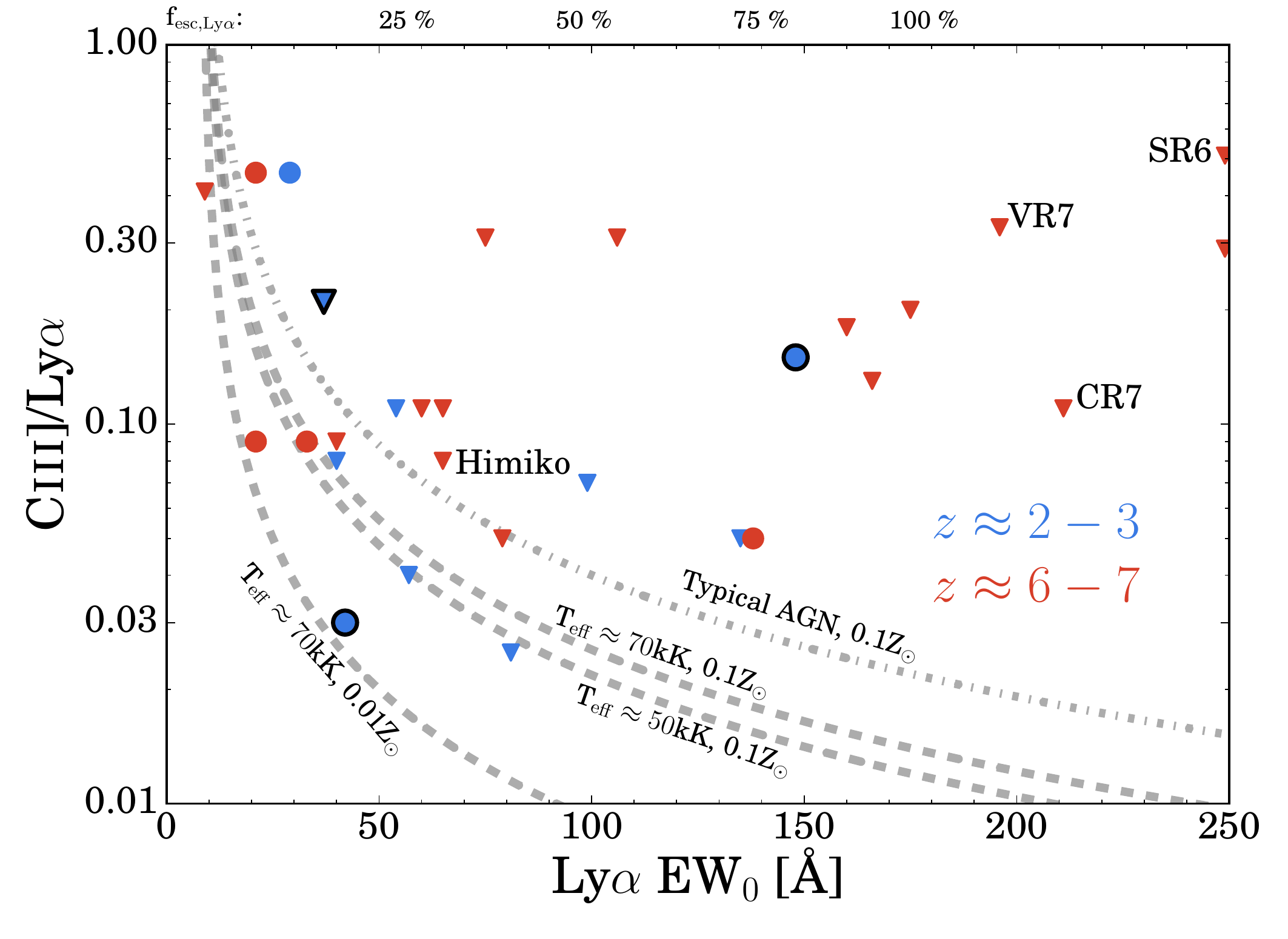}
\end{tabular}
    \caption{Observed C{\sc iv}/Ly$\alpha$ and C{\sc iii}]/Ly$\alpha$ ratios as a function of EW$_{0, \rm Ly\alpha}$ for luminous LAEs at $z\approx2-3$ (Sobral et al. in prep) and the compilation of LAEs and LBGs at $z\approx6-7$ from Table $\ref{tab:compilation}$. Upper limits are shown with down-ward pointing triangles, while detections are shown with circles. We highlight AGN in the $z\approx2-3$ sample with black edges. In dashed (dot-dashed) grey lines, we show expected observed line-ratios based on estimates with the star-forming (AGN) models from Eq. $\ref{eq:model}$. It is clear that C{\sc iv} and C{\sc iii}] are most easily observed at $z\approx6-7$ in galaxies with relatively low Ly$\alpha$ EWs and that the observations of e.g. CR7 and VR7 are not deep enough. }
    \label{fig:C-ratios_EW}
\end{figure*}

Based on Table $\ref{tab:compilation}$, it is already clear that the limits on C{\sc iii}] and C{\sc iv} with respect to Ly$\alpha$ for SR6 and VR7 are higher than, or at most similar to, known detections at $z\sim6-7$, indicating that our observations are not deep enough. As we illustrate in Fig. $\ref{fig:C-ratios}$, we find that the current detections and upper limits at $z\approx6-7$ indicate that C{\sc iv}/Ly$\alpha$ increases towards faint UV luminosities, while it decreases or stays constant at $z\approx2-3$. Contrarily, relatively high C{\sc iii}]/Ly$\alpha$ ratios are detected among UV bright galaxies at $z\approx6-7$, similarly to $z\approx2-3$. In Fig. $\ref{fig:C-ratios_EW}$, we compare the ratios of C{\sc iii}] and C{\sc iv} to Ly$\alpha$ as a function of the Ly$\alpha$ EW$_0$. This illustrates that the $z\approx6-7$ galaxies with observed carbon lines ubiquitously have low Ly$\alpha$ EWs (note that this does not necessarily mean that they are UV bright, as we showed above). This is similar to the comparison sample at $z\approx2-3$ and indicates that the observability of carbon lines may actually be related strongly to the observed strength of Ly$\alpha$ emission and thus on the Ly$\alpha$ escape fraction.

We now compare the measured carbon-Ly$\alpha$ ratios to simple model predictions (Alegre et al. in prep), by correcting for Ly$\alpha$ escape fraction empirically. The physics driving the Ly$\alpha$ escape fraction are complex \citep[e.g.][]{Hayes2015,Henry2015}, with dust, H{\sc i} column density, outflows and (especially at $z>6$) the neutral fraction of the IGM all playing an important role. However, a rough estimate of the Ly$\alpha$ escape fraction may be obtained from the Ly$\alpha$ EW$_0$, as for example shown at $z=2.2$ in \cite{Sobral2016}, see also e.g. \cite{Yang2017} at $z\sim0$. Therefore, we use the EW$_0$ to provide a rough estimate of the Ly$\alpha$ escape fraction, and thus the intrinsic Ly$\alpha$ emission (that is, for example, strongly related to the ionising emissivity). An important caveat here is that the IGM transmission decreases between $z=2.2-6.6$ at wavelengths around Ly$\alpha$ (even into the red wing, \citealt{Laursen2011}), such that the `effective' Ly$\alpha$ escape fraction (including the effect from the IGM) may be under-estimated. On the other hand, this decreasing transmission may be mitigated in the presence of galactic outflows \citep[e.g.][]{Dijkstra2011}. We fit the following relation to the data-points from Fig. 11 (right panel) in \cite{Sobral2016} to estimate the escape fraction: 
\begin{equation}\label{eq:fesc}
f_{\rm esc, Ly\alpha} = 0.006 \rm \,\, EW_{\rm Ly\alpha, 0} - 0.05\,\,\, [5<\rm EW_{\rm Ly\alpha, 0}  < 175],
\end{equation}
where $f_{\rm esc, Ly\alpha}$ is the Ly$\alpha$ escape fraction. Then, we use this relation to estimate observed line-ratios from their theoretically predicted values:
\begin{equation} \label{eq:model}
\frac{f_{\rm CIII]}}{f_{\rm Ly\alpha}} = \frac{\alpha}{f_{\rm esc, Ly\alpha}}\,\,\,\,\, ; \ \,\,\,\, \frac{f_{\rm CIV}}{f_{\rm Ly\alpha}} = \frac{\alpha}{f_{\rm esc, Ly\alpha}},
\end{equation}
where $\alpha$ is the estimated intrinsic line-ratio with respect to Ly$\alpha$. We use the results from {\sc Cloudy} \citep{Ferland2013} modelling (to be presented in Alegre et al. in prep, but with a similar approach to \citealt{Feltre2016}) to model the intrinsic line-ratios. The ionisation sources in these models are either a range of blackbodies (with temperatures ranging from 20 kK to 150 kK, approximating stellar populations and including populations with extreme temperature $>70$kK) and a range of power-laws (with spectral slopes of typical AGN), and the metallicities range from 0.001 Z$_{\odot}$ to solar, see also Sobral et al. in prep. For C{\sc iv}, we use models with values of $\alpha$ between 0.015 (for a blackbody with effective temperature $\approx 70$kK and a metallicity of 0.01 Z$_{\odot}$) and $\alpha=0.11$ (for a typical AGN power-law slope and a metallicity 0.1 Z$_{\odot}$). $\alpha$ decreases rapidly in the case of a lower metallicity or lower effective temperatures. For C{\sc iii}], we use values from $\alpha=0.005$ (T$_{\rm eff}\approx 70$kK, Z$=0.01$ Z$_{\odot}$) to $\alpha=0.022$ for the same AGN model as described above. The results of this modelling is shown in dashed lines in Fig. $\ref{fig:C-ratios_EW}$.

We find that current C{\sc iv} detections at $z\approx6-7$ lie closer to expected line-ratios from AGN than those from star-forming galaxies \citep[c.f.][]{Mainali2017}. However, we note that assuming a higher effective temperature of a stellar population would result in a higher C{\sc iv}/Ly$\alpha$ ratio (compare for example the 60 kK, 0.1 Z$_{\odot}$ line with the 70 kK, 0.1 Z$_{\odot}$ line). Another issue is that Ly$\alpha$ luminosities from these sources are estimated from slits, such that a significant fraction may be missed due to (slightly) more extended emission. On the one hand, the detected C{\sc iii}]/Ly$\alpha$ ratios are already close to those modelled with star-formation as powering source. The C{\sc iv} limits observed in Himiko prefer a star-forming ionising source, or an AGN with a very low metallicity ($\approx 0.01$ Z$_{\odot}$). With the current limits for SR6, CR7 and VR7 this analysis is not very meaningful, and we estimate that we would have to go a factor $5-10$ deeper to detect C{\sc iv} or C{\sc iii}]. 

\begin{figure*}
\begin{tabular}{cc}
	\includegraphics[width=8.65cm]{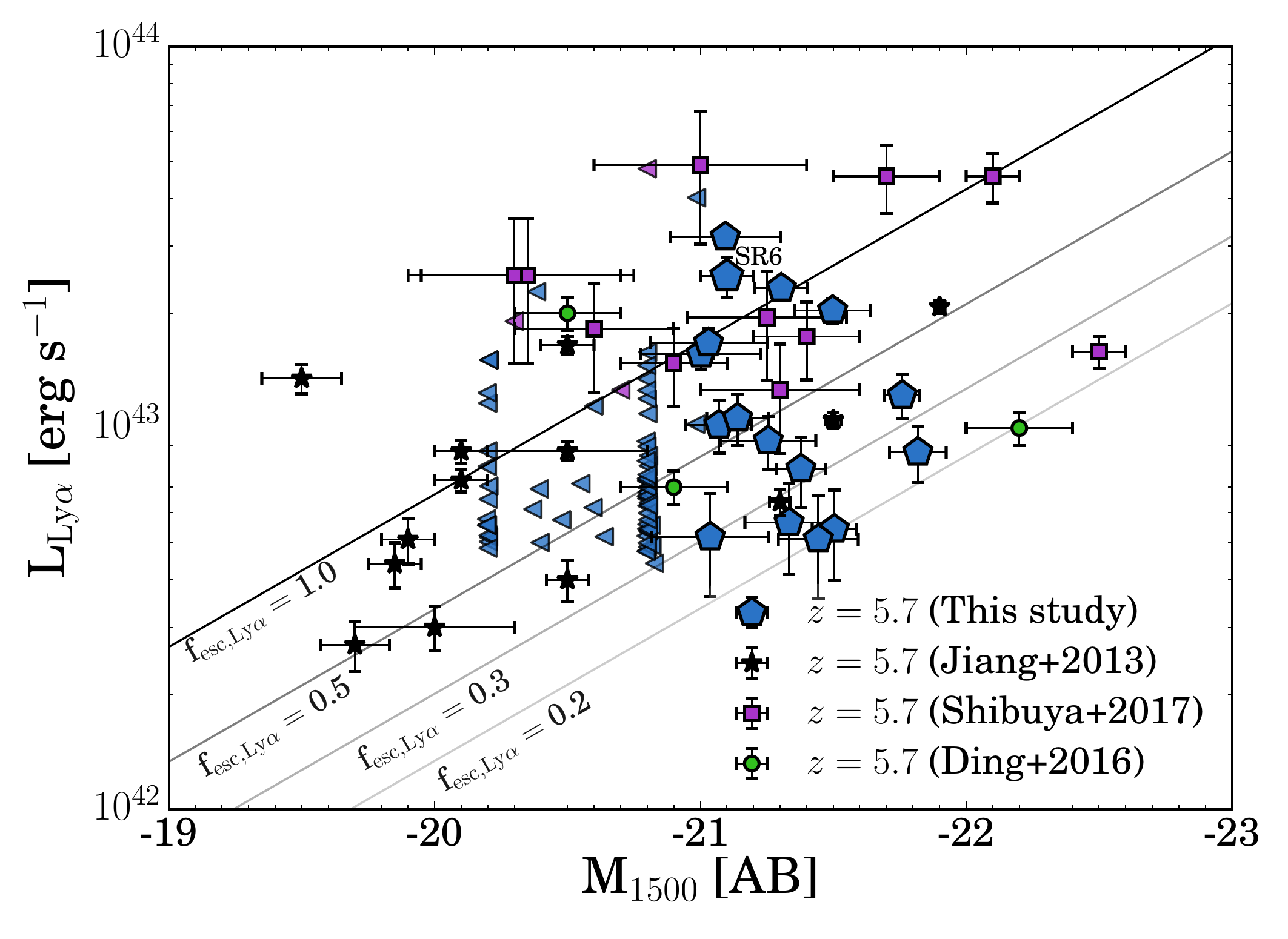}&
	\includegraphics[width=8.65cm]{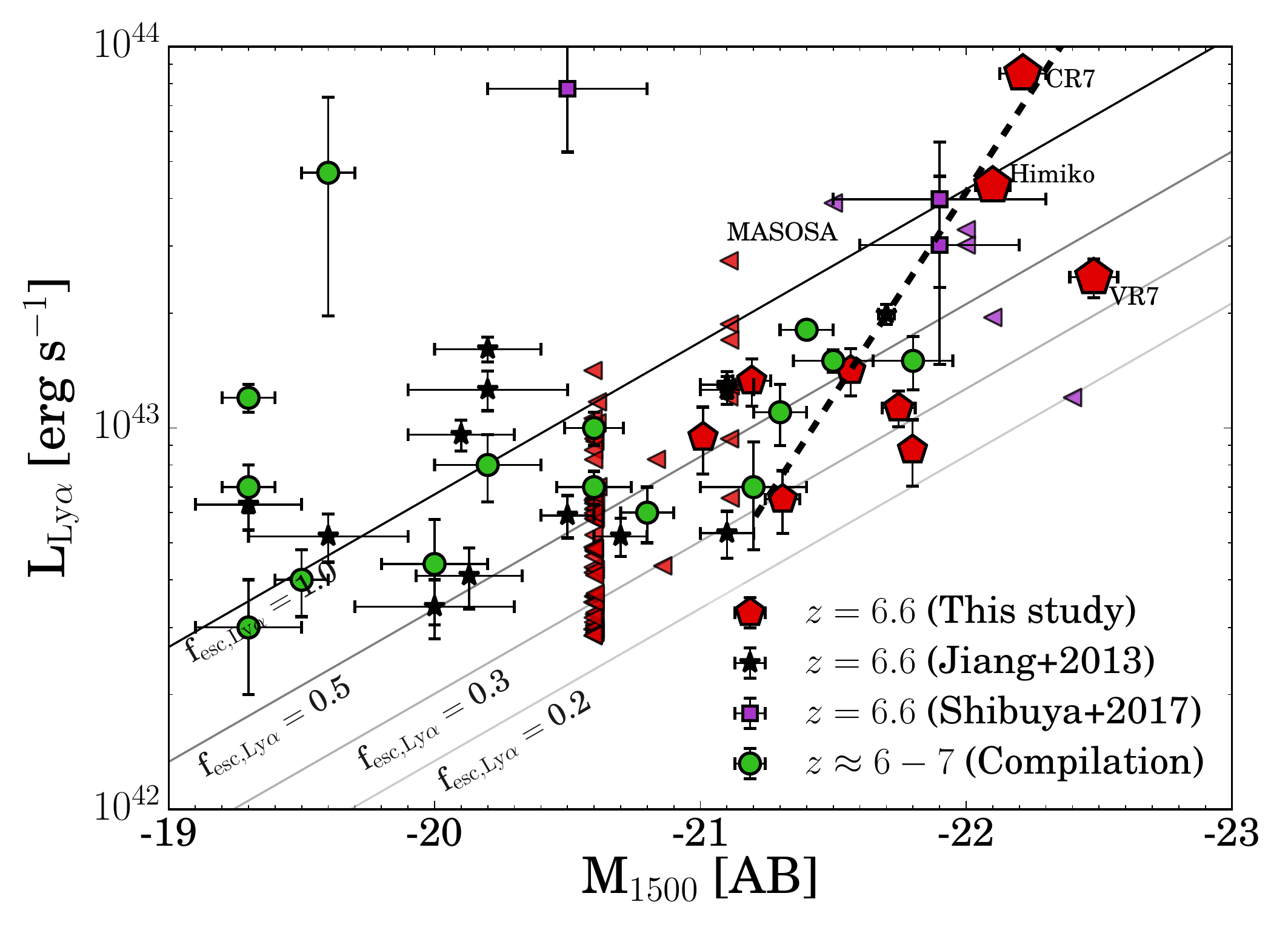}\\
	\end{tabular}
    \caption{M$_{1500}$ versus Ly$\alpha$ luminosity. Rest-frame absolute UV magnitudes are estimated from $Y$ and $J$ band photometry at $z=5.7$ and $z=6.6$, respectively. In the case of a non-detection, we show the local 2$\sigma$ limit. Lines of constant Ly$\alpha$ escape fraction are computed assuming that SFR$_{UV}$=SFR$_{\rm H\alpha}$, case B recombination and no attenuation due to dust. Under these assumptions, several LAEs have escape fractions $>100$ \%. The sources included in this study are the LAEs from \citet{Jiang2013_MUV}, \citet{Matthee2015}, \citet{Santos2016}, \citet{Ding2016} and \citet{Shibuya2017} and the compilation at $z\approx6-7$ comprises the UV-continuum selected galaxies in Table $\ref{tab:references}$ between redshifts $z=6.1$ and $z=7.2$ and with Ly$\alpha$ EW$>20$ {\AA}. In the right panel, the dashed line shows a fit to LAEs with M$_{1500} < -21.2$. Note that Himiko and CR7 have multiple components in the rest-UV. } 
    \label{fig:muv-lya}
\end{figure*}

\subsection{UV luminosities and SFRs of luminous LAEs} \label{sec:uvmags}
In order to investigate how Ly$\alpha$ luminosity is related to the UV luminosity, which traces SFR of timescales of $\sim100$ Myr, we use near-infrared data to measure rest-frame UV luminosities (M$_{1500}$) for LAEs at $z=5.7-6.6$. In addition to the new sources presented in this paper, we also add remaining candidate LAEs at $z=5.7-6.6$ from \cite{Matthee2015} and \cite{Santos2016} and several sources from the literature as described below. 

Rest-frame absolute UV magnitudes of LAEs are estimated from ground-base $Y$ and $J$ band photometry, converted to rest-frame M$_{1500}$ at $z=5.7$ and $z=6.6$ respectively. $Y$ band imaging is available in the UltraVISTA (DR2) coverage of the COSMOS field \citep{McCracken2012}. $J$ band imaging is available in the UDS field through UKIDSS UDS (we use DR8; \citealt{Lawrence2007}), in the COSMOS field through UltraVISTA and in the SA22 field through the UKIDSS DXS. Photometry is measured in 2$''$ apertures with {\sc SExtractor} \citep{Bertin1996} in dual-image mode with the narrow-band image as detection image (i.e. the apertures are centred at the peak Ly$\alpha$ emission). Because the survey depth may vary from source to source (in particular in the COSMOS field due to the UltraVISTA survey design), we measure the depth locally. 2$\sigma$ limits are assigned to sources that are undetected in the NIR imaging. We do not make any corrections for the fact that the effective wavelengths of the filters are not exactly at 1500 {\AA}. However, for a typical UV slope of $\beta\approx-2.3$ \citep[e.g.][]{Ono2010,Jiang2013_MUV}, such a correction would only be on the order of $\Delta$M$_{1500} = 0.004\, (0.03)$ for LAEs at $z=5.7\,(6.6)$. For a more extreme blue or relative red UV slope of $\beta=-3.0$ or $\beta=-1.0$, the correction would be up to 0.1 magnitude. We also add the information from LAEs in the Subaru Deep Field \citep{Kashikawa2011} that have been observed with HST NIR imaging by \cite{Jiang2013_MUV}, LAEs observed by \cite{Ding2016}, recently spectroscopically confirmed LAEs at $z=5.7-6.6$ by \cite{Shibuya2017} and our compilation of UV selected galaxies between $z=6.2-7.2$ with Ly$\alpha$ EW$_0 > 20$ {\AA} (see Table $\ref{tab:compilation}$, green symbols in the right panel of Fig. $\ref{fig:muv-lya}$). In addition to the sources from this compilation, we also added two sources from \cite{Huang2016}, see Table $\ref{tab:references}$. 

Fig. $\ref{fig:muv-lya}$ clearly shows that at fixed Ly$\alpha$ luminosity, there is a large spread in UV luminosities, and vice versa (at both $z=5.7$ and $z=6.6$). Spectroscopically confirmed UV selected galaxies have similar L$_{\rm Ly\alpha}$ luminosities as LAEs. Around L$^{\star}$ (L$_{\rm Ly\alpha} \approx 10^{43}$ erg s$^{-1}$), absolute UV magnitudes can range from up to $3\times$M$_{1500}^{\star}$ (M$_{1500} \approx -21.0$), down to $\approx0.3\times$M$_{1500}^{\star}$, with a 1$\sigma$ spread of 0.9 dex. This means that relatively shallow, wide area Ly$\alpha$ surveys can be an efficient tool to select relatively UV-faint galaxies up to $z\approx7$, thought to be signification contributors to the reionisation process \citep[e.g.][]{Robertson2013,Faisst2016}. It also means that it is challenging to predict Ly$\alpha$ luminosities of UV-continuum selected galaxies, even outside the reionisation epoch.

Fig. $\ref{fig:muv-lya}$ also shows that there is little evidence for a relation between the Ly$\alpha$ luminosity and M$_{1500}$ for Ly$\alpha$ selected sources at $z=5.7$ in our UV and Ly$\alpha$ luminosity range. As both M$_{1500}$ and L$_{\rm Ly\alpha}$ are, to first order, related to the SFR, we would have expected a correlation. To illustrate this, we show lines at constant Ly$\alpha$ escape fractions (based on the assumption that SFR$_{UV}$=SFR$_{\rm H\alpha}$, case B recombination with $T=10 000$K and $n_e = 100$ cm$^{-3}$ and no attenuation due to dust). This result resembles the well known \cite{Ando2006} diagram, which reveals a deficiency of luminous LAEs with bright UV magnitudes between $z\approx5-6$. More recently, other surveys also revealed that the fraction high EW Ly$\alpha$ emitters increases towards fainter UV magnitudes \citep[e.g.][]{Schaerer2011,Stark2011,Cassata2015}. The lack of a strong correlation between M$_{1500}$ and L$_{\rm Ly\alpha}$ might indicate that the SFRs are bursty (because emission-line luminosities trace SFR over a shorter time-scale than UV luminosity), or that the Ly$\alpha$ escape fraction is anti-correlated with M$_{1500}$ (such that Ly$\alpha$ photons can more easily escape from galaxies that are fainter in the UV). A possible explanation for the latter scenario is that slightly more evolved galaxies (which are brighter in the UV) have a slightly higher dust content \citep[e.g.][]{Bouwens2012}, affecting their Ly$\alpha$ luminosity more than the UV luminosity. It is interesting to note that several galaxies lie above the 100 \% Ly$\alpha$ escape fraction line. This implies bursty or stochastic star-formation (which is more likely in lower mass galaxies with faint UV luminosities, e.g. \citealt{MasRibas2016}), alternative Ly$\alpha$ production mechanisms to star-formation (such as cooling), a higher ionising production efficiency (for example due to a top-heavy IMF or binary stars, e.g. \citealt{Gotberg2017}), or dust attenuating Ly$\alpha$ in a different way than the UV continuum \citep[e.g.][]{Neufeld1991,Finkelstein2008,Gronke2016}.

At $z=6.6$, however, current detections indicate a relation between M$_{1500}$ and L$_{\rm Ly\alpha}$, albeit with significant scatter (Fig. $\ref{fig:muv-lya}$). In order to be unbiased due to the depth of $J$ band imaging, we fit a linear relation between log$_{10}$(L$_{\rm Ly\alpha}$) and M$_{1500}$ for LAEs with M$_{1500} < -21.2$ using a least squares algorithm, resulting in:
\begin{equation}
\rm log_{10}(L_{\rm Ly\alpha}/erg \,s^{-1}) = 20.8^{+4.2}_{-4.2} - 1.0^{+0.2}_{-0.2} \, M_{1500} 
 \end{equation}
This fit indicates that for LAEs at $z\approx6.5-7$, M$_{1500}$ and L$_{\rm Ly\alpha}$ are related at 5$\sigma$ significance in the current data (see also \citealt{Jiang2013_MUV}). We measure a (large) 1$\sigma$ scatter of 0.26 dex around this relation. We note that excluding UV selected galaxies results in a lower significance ($\approx3.5\sigma$), but does not significantly change the fit parameters. The fitted slope between M$_{1500}$ and L$_{\rm Ly\alpha}$ is steeper than the slope that is expected at fixed f$_{\rm esc, Ly\alpha}$, which could indicate that the Ly$\alpha$ escape fraction (or its production rate) increases towards brighter magnitudes (for Ly$\alpha$ selected sources). At fainter UV luminosities, we find that the slope is consistent with being flat (within the error-bars) and many of these sources only have upper limits on their UV magnitude. We also note that for a cut at high Ly$\alpha$ luminosity, no clear relation is seen between Ly$\alpha$ and UV luminosity. The presence of very luminous LAEs with luminous UV luminosities at $z=6.6$ is at odds with the \cite{Ando2006} result, indicating additional physical processes playing a role.

We note that the most UV luminous sources at $z=6.6$ have multiple UV-components (CR7, Himiko and VR7), which could help facilitating the escape of Ly$\alpha$ photons. For example, outflows caused by earlier star formation episodes could boost the escape of Ly$\alpha$ photons through the ISM, while the same previous star formation episodes could have ionised a large enough fraction of the IGM around the galaxy such that Ly$\alpha$ can escape. This could be particularly important in the reionisation era at $z\gtrsim6.5$. Similarly, \cite{Jiang2013_MORPH} found that their most UV-luminous LAE at $z=5.7$ and 4 out of the 6 LAEs with M$_{1500} < -20.5$ at $z=6.6$ are interacting/merging. Moreover, the galaxy IOK-1, a confirmed LAE at $z=6.96$ \citep{Iye2006} also consists of two UV-bright components. We note that the spectroscopic follow-up presented in \cite{Furusawa2016} included two luminous UV selected galaxies (M$_{1500} = -22.4, -22.7$) at $z\sim7$ from \cite{Bowler2014}, that have multiple components in the HST imaging \citep{Bowler2017}. These sources are not detected with strong Ly$\alpha$ emission (with a limiting L$_{\rm Ly\alpha} \lesssim 3\times10^{42}$ erg s$^{-1}$), indicating that while multiple components could boost Ly$\alpha$ observability, they do not imply observable Ly$\alpha$ emission at $z\sim7$.

\subsubsection{The production efficiency of ionising photons} \label{sec:xion}
We combine the Ly$\alpha$ and UV measurements to estimate $\xi_{ion}$, the ionising photon production efficiency, which is an important parameter in assessing the ionising budget from star-forming galaxies, particularly in the reionisation era \citep[e.g.][]{Robertson2013,Bouwens2016,MattheeION}. Under the assumption that the escape fraction of ionising photons is close to zero, $\xi_{ion}$ is defined as the number of produced ionising photons per second, per unit UV magnitude:
\begin{equation}
\xi_{ion} =\frac{Q_{ion}}{L_{UV}},
\end{equation}
where $Q_{ion}$ is the number of emitted ionising photons per second, and L$_{UV}$ the UV luminosity at $\lambda_0 \approx 1500$ {\AA}. Ideally, $Q_{ion}$ is estimated from H$\alpha$ measurements, as $L_{\rm H\alpha} = 1.36\times10^{-12} Q_{ion}$ under the assumption that f$_{\rm esc, LyC} = 0$ \% \citep[e.g.][]{Kennicutt1998}. Unfortunately, H$\alpha$ measurements can only be performed at $z>2.5$ after the launch of the {\it James Webb Space Telescope (JWST)}. Therefore, we use the calibration of the Ly$\alpha$ escape fraction with EW$_0$ (see Eq. $\ref{eq:fesc}$), which relates the Ly$\alpha$ luminosity to H$\alpha$ luminosity under the assumption of case B recombination. Rewriting the equations results in:
\begin{equation}
\xi_{ion} =\frac{L_{\rm Ly\alpha}}{8.7\times1.36\times10^{-12}\times f_{\rm esc, Ly\alpha} \times L_{UV}}. 
\end{equation}
Here, the factor 8.7 is the case B recombination ratio between Ly$\alpha$ and H$\alpha$ under typical ISM conditions of T$_e = 10,000$ K and n$_e = 350$ cm$^{-3}$ \citep[e.g.][]{Henry2015}. f$_{\rm esc, Ly\alpha}$ is obtained through Eq. $\ref{eq:fesc}$, with a maximum of 1.0 (for EW$_0 \gtrsim 175$ {\AA}). L$_{UV}$ is computed using the measured M$_{1500}$, assuming negligible dust attenuation (see \citealt{Bouwens2016} for a discussion on how dust attenuation affects $\xi_{ion}$). This empirically motivated method to estimate $\xi_{ion}$ can easily be tested with follow-up observations with {\it JWST}.

\begin{figure}
	\includegraphics[width=8.65cm]{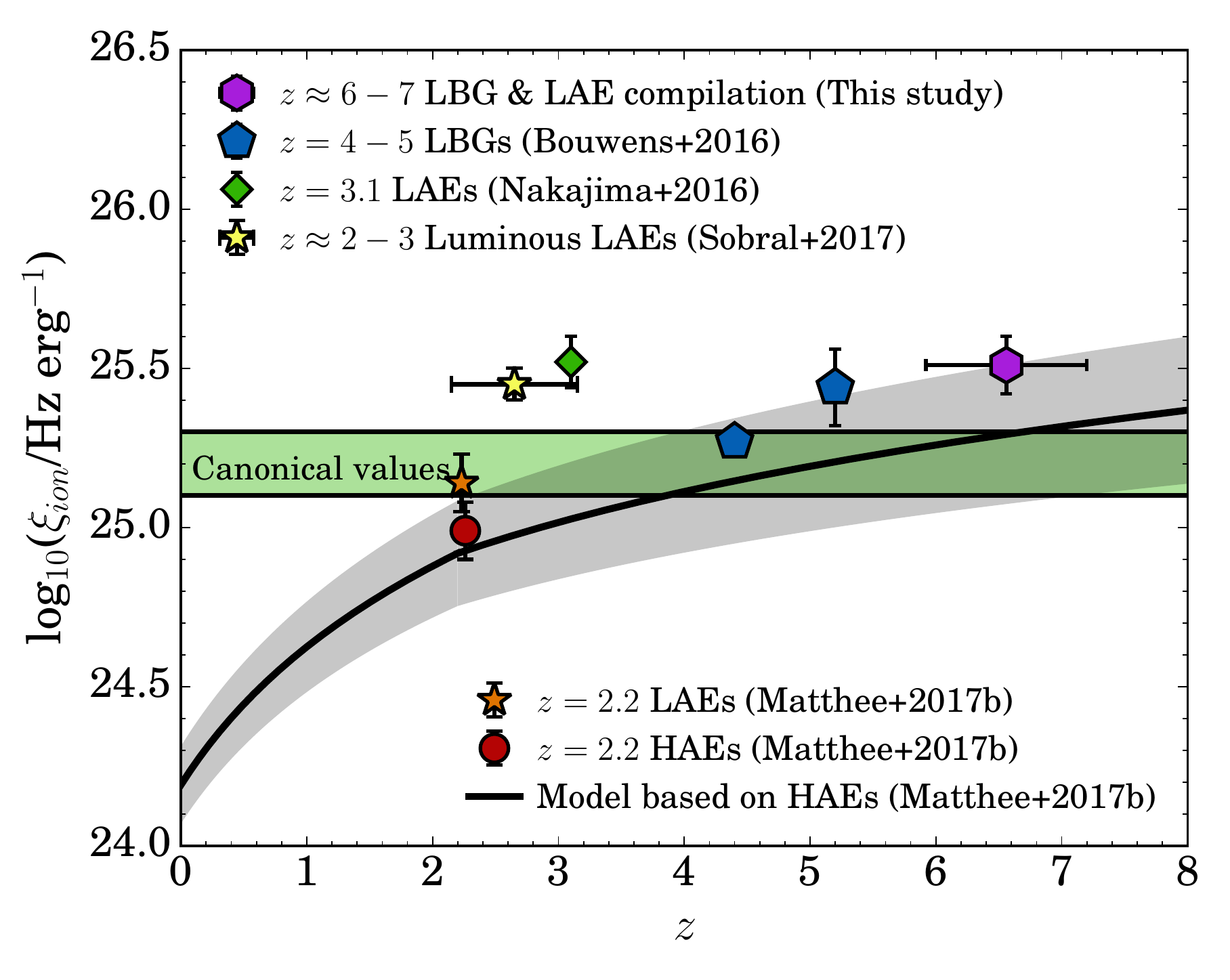}
    \caption{Production efficiency of ionising photons ($\xi_{ion}$) versus redshift for different compilations of galaxies, showing our results at $z\approx6-7$ with a purple hexagon. The green region shows the canonical values from \citet{Robertson2013}. The black line shows the modelled/predicted evolution of $\xi_{ion}$ from \citet{MattheeION} based on the redshift evolution of the H$\alpha$ EW and the relation between $\xi_{ion}$ and H$\alpha$ EW for H$\alpha$ emitters (HAEs). It can be seen that at $z\approx2-3$ LAEs have a significantly higher $\xi_{ion}$ than the typical HAE/star-forming galaxy. Samples of LBGs at $z\approx4-5$ and the compilation of LBGs and LAEs from Table $\ref{tab:compilation}$ at $z\approx6-7$ show elevated values of $\xi_{ion}$ compared to the canonical value, and qualitatively confirm the trend of the modeled evolution of $\xi_{ion}$. } 
    \label{fig:xion}
\end{figure}

Using this prescription, we calculate values of log$_{10}(\xi_{ion}$/Hz erg$^{-1}$) $\gtrsim 25.25\pm0.23$ and $\gtrsim 24.66\pm0.17$ for SR6 and VR7, respectively. We write these as upper limits because the Ly$\alpha$ EWs of SR6 and VR7 indicate f$_{\rm esc, Ly\alpha} = 100$ \%, which may be an over-estimate (in the case of 50 \% Ly$\alpha$ escape, $\xi_{ion}$ would increase by 0.3 dex). For CR7 and Himiko we measure log$_{10}(\xi_{ion}$/Hz erg$^{-1}$) $\gtrsim 25.33\pm0.06$ and log$_{10}(\xi_{ion}$/Hz erg$^{-1})=25.55\pm0.07$, respectively. Except for VR7, these values are similar to the measurements of faint LAEs at $z\approx3$ \citep{Nakajima2016} and $z\approx4-5$ Lyman-break galaxies \citep{Bouwens2016}. The median value of $\xi_{ion}$ for our compilation in Table $\ref{tab:compilation}$ is log$_{10}(\xi_{ion}$/Hz erg$^{-1}) = 25.51\pm0.09$ (Fig. $\ref{fig:xion}$), which is similar to the mean value of the reference sample of luminous LAEs at $z\approx2-3$, for which we measure a mean log$_{10}(\xi_{ion}$/Hz erg$^{-1}) = 25.45\pm0.05$ using the same method, and slightly lower than those measured by \cite{Schaerer2016} in a sample of low redshift Lyman-Continuum leakers.

We also compare our method to the values of $\xi_{ion}$ obtained independently by \cite{Stark2015_CIV,Stark2016} using photo-ionisation modelling of UV metal lines. While we measure a higher value of log$_{10}(\xi_{ion}$/Hz erg$^{-1}) = 26.63\pm0.36$ for EGS-zs8-2, our results for A1703\_zd6, EGS-zs-1 (see Table $\ref{tab:compilation}$) and COS zs7-1 are log$_{10}(\xi_{ion}$/Hz erg$^{-1}) = 25.52\pm0.18, 25.64\pm0.12$ and $25.63\pm0.15$, encouragingly consistent with the estimates from Stark et al. Under the model-assumptions, these results indicate that luminous Ly$\alpha$ emitters produce ionising photons a factor two more efficiently than typically assumed \citep[e.g.][]{Robertson2013}, and similar to the estimated $\xi_{ion}$ based on the evolution of the H$\alpha$ EW / specific SFR \citep{MattheeION}, see Fig. $\ref{fig:xion}$. This implies that a significant amount of photons that reionised the Universe may have been produced in LAEs, in particular if the ISM conditions in these galaxies are also facilitating LyC photons to escape \citep[e.g.][]{Dijkstra2016}, which can be constrained with future observations with {\it JWST} \citep[e.g.][]{Zackrisson2017}.

\section{Conclusions} \label{sec:conclusions} 
We have presented the results of X-SHOOTER follow-up observations of a sample of luminous LAE candidates at $z=5.7-6.6$ in the SA22 field. We present the properties of newly confirmed LAEs (summarised in Table $\ref{tab:properties}$) and compare them with other LAEs and LBGs at similar redshifts. The main results are:

\begin{enumerate}
\item We spectroscopically confirm SR6, the most luminous LAE at $z=5.676$ in the SA22 field. While SR6 has a high Ly$\alpha$ luminosity and extreme EW (L$_{\rm Ly\alpha} = 2.5\pm0.3\times10^{43}$ erg s$^{-1}$, EW$_0 > 250$ {\AA}), it has a typical UV continuum luminosity (M$_{1500}= -21.1\pm0.1$) and a narrow Ly$\alpha$ line (236$\pm16$ km s$^{-1}$).

\item We confirm VR7, the most luminous LAE at $z=6.532$, in the SA22 field (L$_{\rm Ly\alpha} = 2.4\pm0.2\times10^{43}$ erg s$^{-1}$, EW$_0 > 196$ {\AA} and $v_{\rm FWHM} = 340\pm14$ km s$^{-1}$, see \S \ref{sec:VR7}). Among the luminous LAEs known at $z\sim6.5$, VR7 is also the most luminous in the UV found so far (M$_{1500} = -22.5\pm0.2$). Despite this luminosity, we do not detect any signs of AGN activity (such as C{\sc iv} or Mg{\sc ii} emission) in the spectrum at the current depths ($f\lesssim 2\times10^{-17}$ erg s$^{-1}$ cm$^{-2}$, EW $\lesssim 20$ {\AA}). In contrast, essentially all LAEs at $z\approx2-3$ with similar Ly$\alpha$ and UV luminosities are AGN.

\item Ly$\alpha$ line-widths increase slowly with Ly$\alpha$ luminosity at $z=5.7$, while such a trend is not seen at $z=6.6$. We find indications that the line-widths of LAEs with L$_{\rm Ly\alpha} \approx 10^{42.5}$ erg s$^{-1}$ increase between $z=5.7-6.6$ (\S \ref{sec:widths}), although at relatively low statistical significance due to small sample sizes. This evolution occurs at the same luminosity where the number densities decrease, and Ly$\alpha$ spatial scales increase \citep{Santos2016}, all indicating patchy reionisation. 

\item In \S $\ref{sec:compilation}$, we argue empirically that rest-UV lines besides Ly$\alpha$ are most easily observed in galaxies with relatively low Ly$\alpha$ EWs. This explains why carbon lines have been detected in luminous UV-continuum selected sources, while they have not been easily detected in Ly$\alpha$ selected sources. 

\item Combining our sources with a compilation of LAEs and LBGs at $z\approx6-7$, we do not detect a clear relation between the Ly$\alpha$ luminosity and absolute UV magnitude at $z=5.7$ indicating a lower Ly$\alpha$ escape fraction at brighter UV luminosities, for example due to dust (\S \ref{sec:uvmags}). There is a large dispersion in absolute UV magnitudes of $>L^{\star}$ LAEs of $\sigma=0.9$ dex.  

\item At $z=6.6$, we find that, at M$_{1500} < -21$, the Ly$\alpha$ and UV luminosity are strongly correlated, while there is no evidence for a relation at fainter UV luminosities. This means that the Ly$\alpha$ escape fraction and/or its production rate increases strongly among luminous LAEs between $z=5.7-6.6$. Most luminous LAEs show multiple components in the rest-UV. This could indicate that such merging systems could boost effective Ly$\alpha$ transmission through the IGM at $z>6.5$, increasing the effective Ly$\alpha$ escape fraction.

\item Under basic assumptions, we find that several LAEs at $z\approx6-7$ would have Ly$\alpha$ escape fractions $\gtrsim100$ \%, which could indicate bursty star-formation histories, alternative Ly$\alpha$ production mechanisms, a higher ionising production efficiency, or dust attenuating Ly$\alpha$ in a different way than the UV continuum.
 
\item Using an empirical relation to estimate the Ly$\alpha$ escape fraction, we present a method to compute $\xi_{ion}$, the production efficiency of ionising photons, based on Ly$\alpha$ and UV continuum measurements (\S $\ref{sec:xion}$). Our results indicate that luminous LAEs at $z\approx6-7$ produce ionising photons efficiently, with a median log$_{10}(\xi_{ion}$/Hz erg$^{-1}) = 25.51\pm0.09$, similar to other recent measurements of LAEs and LBGs at $z\approx2-5$. These measurements will easily be testable with {\it JWST}.
 
\end{enumerate}

In the future, significant improvements can be made by observing a statistical sample of homogeneously selected Ly$\alpha$ emitters at $z=5.7-6.6$ with IFU spectroscopy with {\it JWST}, which can measure H$\alpha$ up to $z=6.6$. Such measurements can constrain any evolution in the effective Ly$\alpha$ escape fraction directly (due to an increasingly neutral IGM), by controlling for the apertures/spatial scales and positions of the emission (hence the IFU), and allow us to test the empirical models to estimate Ly$\alpha$ escape fractions and $\xi_{ion}$. The bright, spectroscopically confirmed LAEs are the ideal targets to pioneer such studies as they already show extended/multiple component morphologies.

\section*{Acknowledgments}
We thank the referee for a constructive report that has improved the quality and clarity of this work. The authors thank Grecco Oyarz\'un for discussions. JM acknowledges the support of a Huygens PhD fellowship from Leiden University. DS acknowledges financial support from the Netherlands Organisation for Scientific research (NWO) through a Veni fellowship and from Lancaster University through an Early Career Internal Grant A100679. BD acknowledges financial support from NASA through the Astrophysics Data Analysis Program (ADAP), grant number NNX12AE20G. We thank Kasper Schmidt for providing measurements. Based on observations with the W.M. Keck Observatory through program C267D. The W.M. Keck Observatory is operated as a scientific partnership among the California Institute of Technology, the University of California and the National Aeronautics and Space Administration. Based on observations made with ESO Telescopes at the La Silla Paranal Observatory under programme IDs 097.A-0943, 294.A-5018 and 098.A-0819 and on data products produced by TERAPIX and the Cambridge Astronomy Survey Unit on behalf of the UltraVISTA consortium. The authors acknowledge the award of observing time (W16AN004) and of service time (SW2014b20) on the William Herschel Telescope (WHT). WHT and its service programme are operated on the island of La Palma by the Isaac Newton Group in the Spanish Observatorio del Roque de los Muchachos of the Instituto de Astrofisica de Canarias.
Based on observations made with the NASA/ESA Hubble Space Telescope, obtained [from the Data Archive] at the Space Telescope Science Institute, which is operated by the Association of Universities for Research in Astronomy, Inc., under NASA contract NAS 5-26555. These observations are associated with program \#14699.
We are grateful for the excellent data-sets from the COSMOS, UltraVISTA, SXDS, UDS and CFHTLS survey teams, without these legacy surveys, this research would have been impossible. We have benefited from the public available programming language {\sc Python}, including the {\sc numpy, matplotlib, pyfits, scipy} and {\sc astropy} packages, the astronomical imaging tools {\sc SExtractor, Swarp} and {\sc Scamp} and the {\sc Topcat} analysis tool \citep{Topcat}.




\bibliographystyle{mnras}

\bibliography{bib_LAEevo.bib}




\appendix
\section{Galaxy compilation} \label{sec:compilation_info}
Here we describe shortly the details of the sample of sources listed in Table $\ref{tab:references}$. Part of this sample are narrow-band selected LAEs, where the Ly$\alpha$ flux is measured with a NB (except for WISP302, which is measured with the HST/WFC3 grism, \citealt{Bagley2017}). The other part of the sample are UV selected Lyman-break galaxies for which the Ly$\alpha$ flux and EW have been measured from slit spectroscopy, except for the grism measurements of RXCJ2248.7-4431 \citep{Schmidt2017}, the GLASS-stack \citep{Schmidt2016} and FIGS\_GN1\_1292 \citep{Tilvi2016}. We note that due to extended Ly$\alpha$ emission and slit losses their Ly$\alpha$ luminosities may be under-estimated, in particular when Ly$\alpha$ is offset from the UV emission \citep[e.g.][]{Vanzella2017}. For CR7, we use updated constraints on metal-lines from the recalibrated spectrum that will be presented in Sobral et al. in prep.

In the case Ly$\alpha$ luminosities are not published, we have computed them from the published line-flux and luminosity distance corresponding to the source redshift. For sources from \cite{Ding2016}, we have used luminosities from their discovery-papers (\citealt{Ouchi2005} and \citealt{Shimasaku2006}). If not published, M$_{1500}$ is computed based on the observed magnitude in the band closest to a rest-frame $\lambda = 1500$ {\AA}, corrected for the distance modulus and band-width spreading and for possible lensing magnification. In the case of emission-line doublets (such as C{\sc iii}]$_{1907,1909}$), we use the sum of both lines, except in the case of the O{\sc iii}] doublet of A1703\_zd6, where only one component is measured due to adjacent skylines. Most measurements of/limits on rest-UV lines besides Ly$\alpha$ and N{\sc v} have been performed with slit spectroscopy from the ground and are thus significantly hampered by the sky OH emission lines in the near-infrared. This is less of an issue for grism data, although these are limited by their spectral resolution, and thus mostly sensitive to high EW lines. Another employed technique used a matched narrow-band that measures specific emission-lines at specific redshifts, for example C{\sc iii}] at $z=5.7$ \citep{Cai2011,Cai2015,Ding2016}.

\begin{table}
\centering
\caption{Galaxies included in compilations of line-widths and rest-UV line-ratios.}
\begin{tabular}{lrp{3cm}}
\hline
ID & Redshift & Reference \\ \hline
\bf Ly$\alpha$ selected & &\\
SGP 8884 & 5.65 & \cite{Westra2006,Lidman2012}\\ 
SR6 & 5.676 & \bf This paper\\
Ding-3 &  5.69 & \cite{Ouchi2005,Ding2016} \\  
Ding-4 & 5.69 & \cite{Ouchi2005,Ding2016} \\  
Ding-5 & 5.69 & \cite{Ouchi2005,Ding2016} \\
Ding-2 &  5.692 & \cite{Ouchi2005,Ding2016}  \\  
Ding-1 &  5.70  & \cite{Shimasaku2006,Ding2016}  \\   
J233408 & 5.707 & \cite{Shibuya2017} \\
S11 5236 & 5.72 & \cite{Westra2006,Lidman2012}\\ 
J233454 &  5.732 & \cite{Shibuya2017} \\
J021835 & 5.757 & \cite{Shibuya2017} \\
WISP302 & 6.44 & \citet{Bagley2017} \\
VR7 & 6.532 & \bf This paper\\
LAE SDF-LEW-1 & 6.538 & \cite{Kashikawa2012} \\ 
J162126 &  6.545 & \cite{Shibuya2017} \\
J160940 & 6.564 & \cite{Shibuya2017} \\
J100550 & 6.573 & \cite{Shibuya2017} \\
J160234 &  6.576& \cite{Shibuya2017} \\
Himiko  & 6.59 & \cite{Ouchi2009,Zabl2015}\\ 
COLA1 & 6.593 & \cite{Hu2016} \\
CR7 & 6.604 & \cite{Sobral2015} \\

 & & \\

\bf UV selected & &\\
WMH S & 5.618 & \citet{Willott2013} \\ 
WMH 13 & 5.983 & \citet{Willott2013} \\ 
A383-5.2  & 6.0294 & \cite{Richard2011,Stark2015_CIII} \\ 
WMH 5 & 6.068 & \citet{Willott2013} \\ 
RXCJ2248.7-4431-ID3 & 6.11 & \cite{Mainali2017} \\ 
RXCJ2248.7-4431 &  6.11 & \cite{Schmidt2017} \\ 
CLM 1 & 6.17 & \citet{Cuby2003} \\ 
MACS0454-1251 & 6.32 & \citet{Huang2016} \\
RXJ1347-1216 & 6.76 & \citet{Huang2016} \\ 
SDF-46975& 6.844 & \cite{Ono2012} \\ 
IOK-1 & 6.96 & \cite{Iye2006,Cai2011,Ono2012} \\ 
BDF-521 & 7.01 &  \cite{Vanzella2011,Cai2015}  \\ 
A1703\_zd6  & 7.045 & \cite{Stark2015_CIV} \\ 
BDF-3299 & 7.109 & \cite{Vanzella2011} \\ 
GLASS-stack & $<7.2>$ & \cite{Schmidt2016} \\  
GN-108036 & 7.213 & \cite{Ono2012,Stark2015_CIII} \\  
EGS-zs8-2  & 7.477 & \cite{Stark2016} \\ 
FIGS\_GN1\_1292 & 7.51 & \cite{Finkelstein2013,Tilvi2016} \\ 
EGS-zs8-1 & 7.73 & \cite{Oesch2015,Stark2016} \\ 

\hline\end{tabular}
\label{tab:references}
\end{table}

\section{Catalogues of candidate LAEs at $z=5.7-6.6$} \label{sec:catalogs}
We publish catalogues of all candidate LAEs at $z=5.7$ from \cite{Santos2016} and at $z=6.6$ from \cite{Matthee2015} with the paper. The first five entries of these catalogues are shown in Table $\ref{tab:z6}$ and $z=6.6$ Table $\ref{tab:z7}$.

\begin{table*}
\caption{First five entries of our candidate LAEs at $z=5.7$ from \citet{Santos2016}. Full electronic table is available online. Line-flux (f$_{\rm NB816}$), Ly$\alpha$ luminosity (assuming a luminosity distance corresponding to $z=5.7$) and EW$_0$ are measured in 2$''$ apertures. For sources with spectroscopically confirmed redshift, we corrected the luminosity for the narrow-band filter transmission at the wavelength where the line is observed. }
\begin{tabular}{|l|l|l|l|l|l|}
\hline
  \multicolumn{1}{|c|}{ID} &
  \multicolumn{1}{c|}{R.A.} &
  \multicolumn{1}{c|}{Dec.} &
  \multicolumn{1}{c|}{f$_{\rm NB816}$} &
  \multicolumn{1}{c|}{L$_{\rm Ly\alpha}$} &
  \multicolumn{1}{c|}{EW$_0$} \\
& J2000 & J2000 & $10^{-17}$ erg s$^{-1}$ cm$^{-2}$& $10^{42}$ erg s$^{-1}$ & {\AA} \\ \hline
   SA22-NB816-480736 & 22:17:28.81 & +00:53:02.29 & 6.6 & 24.6 & 178\\
   SA22-NB816-444574 & 22:17:29.41 & +00:34:13.07 & 2.3 & 8.5 & 27\\
   SA22-NB816-429880 & 22:17:33.65 & +00:26:47.99 & 1.4 & 5.1 & 25\\
   SA22-NB816-429969 & 22:17:36.16 & +00:26:49.65 & 3.7 & 13.6 & 380\\
   SA22-NB816-438282 & 22:17:37.24 & +00:30:57.20 & 3.9 & 14.3 & 43\\

\hline\end{tabular}
\label{tab:z6}
\end{table*}

\begin{table*}
\caption{First five entries of our candidate LAEs at $z=6.6$ from \citet{Matthee2015}. Full electronic table is available online. Line-flux (f$_{\rm NB921}$), Ly$\alpha$ luminosity (assuming a luminosity distance corresponding to $z=6.55$) and EW$_0$ are measured in 2$''$ apertures. For sources with spectroscopically confirmed redshift, we corrected the luminosity for the narrow-band filter transmission at the wavelength where the line is observed. }
\begin{tabular}{|l|l|l|l|l|l|}
\hline
  \multicolumn{1}{|c|}{ID} &
  \multicolumn{1}{c|}{R.A.} &
  \multicolumn{1}{c|}{Dec.} &
  \multicolumn{1}{c|}{f$_{\rm NB921}$} &
  \multicolumn{1}{c|}{L$_{\rm Ly\alpha}$} &
  \multicolumn{1}{c|}{EW$_0$} \\
& J2000 & J2000 & $10^{-17}$ erg s$^{-1}$ cm$^{-2}$& $10^{42}$ erg s$^{-1}$ & {\AA} \\ \hline
 VR7 & 22:18:56.36 & +00:08:07.32 & 4.8 & 23.4 & 203\\
  COSMOS-NB921-20802 & 10:02:07.83 & +02:32:17.25 & 1.4 & 7.0 & 56\\
  COSMOS-NB921-5032 & 10:02:04.33 & +02:20:29.12 & 2.4 & 11.7 & 125\\
  COSMOS-NB921-107681 & 10:01:54.68 & +02:10:18.59 & 2.1 & 10.3 & 100\\
  COSMOS-NB921-100684 & 10:01:27.54 & +02:06:46.47 & 3.5 & 17.0 & 116\\
\hline\end{tabular}
\label{tab:z7}
\end{table*}

\bsp	
\label{lastpage}
\end{document}